%% file: main-clean.tex
\documentclass[%
reprint,
superscriptaddress,
longbibliography,
noeprint,nofootinbib,
amsmath,amssymb,
aps
]{revtex4-2}
\usepackage[utf8]{inputenc}
\usepackage{bbm}
\usepackage{braket}
\usepackage{bm}
\newcommand{\ec}[1]{\ensuremath{\varepsilon_{\rm {#1}}}}
\usepackage{enumitem}
\usepackage{amsmath,amsfonts,amssymb,amsthm,graphics,graphicx, physics}

\usepackage{chngcntr}

\theoremstyle{definition}
\usepackage [autostyle, english = american]{csquotes}
\usepackage[ruled,vlined]{algorithm2e}
\SetKwFunction{MCSweep}{MCSweep}
\SetKwFunction{MCSweepM}{MCSweepM}
\SetKwProg{Fn}{Function}{:}{}
\graphicspath{{figures/}}
\MakeOuterQuote{"}
\usepackage{graphicx} 
\usepackage[colorlinks=true,linkcolor=blue,allcolors=blue]{hyperref}
\usepackage[capitalize,nameinlink]{cleveref}

\def\equationautorefname~#1\null{Eq. (#1)\null}
\newcommand{\appref}[1]{\hyperref[#1]{App.~\ref*{#1}}}
\usepackage{subfiles}
\newcommand{\comment}[1]{}
\renewcommand{\vec}{\bm}
\newcommand{\Q}[1]{Q_{\beta,\alpha,\gamma}^{#1}}
\newcommand{\tQ}[2]{\tilde{Q}_{{#1},\beta,\alpha,\gamma}^{#2}}
\newcommand{\spin}{\sigma}
\newcommand{\sconf}{\ensuremath{s_{\rm conf}}}

\newcommand{\Eglass}{\ensuremath{\epsilon_{\rm G}}}
\newcommand{\nL}{\ensuremath{n_{\rm L}}}
\newcommand{\kL}{\ensuremath{k_{\rm L}}}
\newcommand{\Tdyn}{\ensuremath{T_{\rm mem}}}
\newcommand{\Tglass}{\ensuremath{T_{\rm G}}}

\makeatletter 
    
\renewcommand\onecolumngrid{
\do@columngrid{one}{\@ne}%
\def\set@footnotewidth{\onecolumngrid}
\def\footnoterule{\kern-6pt\hrule width 1.5in\kern6pt}%
}

\renewcommand\twocolumngrid{
        \def\footnoterule{
        \dimen@\skip\footins\divide\dimen@\thr@@
        \kern-\dimen@\hrule width.5in\kern\dimen@}
        \do@columngrid{mlt}{\tw@}
}%

\makeatother    
\begin{document}
\subfile{main-prl-clean}

\input{main.bbl}
\clearpage
\onecolumngrid
\setcounter{figure}{0}
\setcounter{section}{0}
\setcounter{page}{1}
\let\oldthefigure\thefigure
\renewcommand{\thefigure}{S\oldthefigure}
\setcounter{equation}{0}
\counterwithout{equation}{section}
\renewcommand{\theequation}{S\arabic{equation}}
\newpage
\subfile{supp-clean}
\end{document}

%% file: main-prl-clean.tex
\title{Stable valleys in the glassy landscape of a low-density parity-check (LDPC) code}
\author{Grace M. Sommers}
\email{grace.sommers@yale.edu}
\affiliation{Department of Physics, Princeton University, Princeton, NJ 08544, USA}

\author{David A. Huse}
\affiliation{Department of Physics, Princeton University, Princeton, NJ 08544, USA}
\begin{abstract}
Classical low-density parity-check (cLDPC) codes defined on expander graphs are a fundamental ingredient in the construction of good quantum LDPC codes, a recent milestone in quantum error correction. They also define interesting statistical mechanics models in their own right, as they include examples of spin glass order without quenched randomness or frustration. We investigate this via a case study of a cLDPC code on a locally tree-like expander graph. Recursive techniques on trees, made possible by the locally tree-like property, probe a menagerie of stable, incongruent valleys induced by imposing different boundary conditions at low temperature. Analysis of stable and unstable fixed points within the cavity method, complemented by a numerical study of the valleys on closed finite graphs, reveals the inequivalence of the microcanonical and canonical ensemble for certain valleys.

\end{abstract}
\maketitle 
Glassiness -- a rugged free energy landscape resulting in ergodicity breaking without conventional symmetry breaking -- arises in both amorphous materials (structural glasses)~\cite{debenedetti_stillinger_supercooled2001, tarjus2005frustration, Sastry2006,bertier2011rmp,Stillinger2013} and magnetically disordered systems (spin glasses)~\cite{Edwards1975,anderson1978concept,sherrington1975solvable,binder1986review,kirkpatrick1987p,Dahlberg2025,Tahiri2026}. While the latter are often characterized by quenched disorder and frustration~\cite{Toulouse1977,anderson1978concept,Dahlberg2025,Tahiri2026}, neither of these features is an essential ingredient~\cite{Marinari_1994a,Marinari_1994,bouchaud1994self,Franz1995,Cugliandolo1995,Lipowski2000,Yoshino2018,Dong2021,niedda2023}. Non-random and/or unfrustrated spin glasses with finite connectivity have long been studied in the context of satisfiability~\cite{franz2001ferromagnet,Ricci2001,mezard2003two_solutions,montanari2003nature,montanari2004cooling_schedule,Bellitti2021,Bernaschi2021} and classical error correction~\cite{gallager1960thesis,gallager1962low,montanari2001glassy,franz2002dynamic,mezard2009information},
but only recently has the mechanism behind spin-glass order in such models been made rigorous, with Ref.~\cite{Placke_glass} proving the existence of a low-temperature spin-glass phase in low-density parity-check models with exponentially many ground states surrounded by extensive energy barriers.

Extensive energy barriers arise naturally when the Hamiltonian is defined on an expander graph, defined below. However, there exist many models 
for which extensive barriers have not been proven, but which nevertheless empirically exhibit hallmarks of spin glass order~\cite{Placke_glass}. The objective of the present work is therefore to construct a detailed, quantitative phase diagram for one such model. Employing a set of complementary analytical and numerical methods, we uncover a rich phenomenology of different instabilities. We also find suggestive evidence for an intermediate regime with valleys that are microcanonically stable yet canonically unstable, revealing the inequivalence of ensembles for models with extensive energy barriers.

\emph{Model.---}
\begin{figure}[t]
\includegraphics[width=\linewidth]{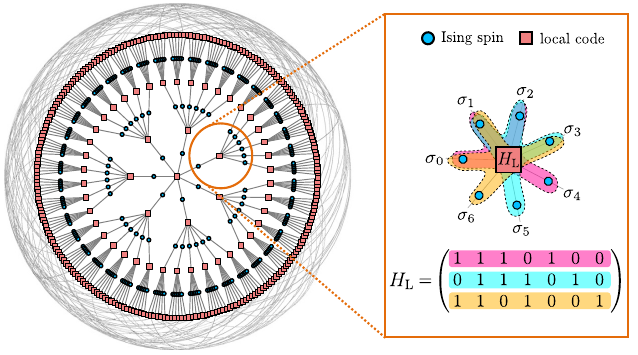}
\caption{The LDPC code studied in this work is defined on a high-girth random regular graph, where the external thin gray lines schematically indicate randomly closed boundaries. Each vertex (orange square) is satisfied if the neighboring bits (blue circles) belong to the code space of the chosen local code, with parity check matrix $H_{\rm L}$. Here we take the local code to be the Hamming [7,4,3] code. (Adapted from Fig. 5 of Ref.~\cite{Placke_glass}.) \label{fig:model-LDPC}}
\end{figure}
Our Hamiltonian of interest belongs to a class of models known as \textit{classical low-density parity-check (cLDPC) codes}: each spin is involved in finitely many interactions, and each interaction involves a finite number of spins~\cite{guruswami2019essential}. To define the model, consider a \textit{high-girth random regular graph} (HGRRG)~\cite{rrg} with a two-state classical (Ising) spin $\sigma$ placed on each edge (\autoref{fig:model-LDPC}). Spins interact only if they are incident to a common vertex. The interaction on each vertex $v$ divides the vertices into two sets: either the vertex is ``satisfied'', and the energy of that vertex is $E_v(\vec{\sigma}_v) = 0$ (its ground state), or the vertex is ``violated'', and $E_v(\vec{\sigma}_v)=+1$. The local spin configurations satisfying the vertex are codewords of a linear binary code, $\mathcal{C}$. The same code is used deterministically at every vertex~\cite{perms}.

The global ground states are spin configurations which satisfy every vertex. 
Suppose the local code $\mathcal{C}$ is a $[\nL, \kL, d_{\rm L}]$ classical linear code: its codewords are configurations of the $\nL$ spins adjacent to a vertex, satisfying $(\nL-\kL)$ independent constraints, such that two codewords differ on at least $d_{\rm L}$ spins. The global code, known as a \textit{Tanner code}~\cite{tanner1981recursive}, defined on a graph with $N_v$ vertices and $N_v \nL / 2$ edges, has at most $(\nL-\kL) N_v$ independent constraints (``checks''), and thus for $\kL > \nL/2$ there is an exponentially large ground space, containing $2^K$ configurations with $K \geq N_v (\kL - \nL/2)$. For our model of interest, introduced in Ref.~\cite{Placke_glass}, this inequality is empirically saturated: there are no \textit{redundancies} between the checks on different vertices~\cite{Placke_glass,Hong2025}.

Lack of redundancies implies the exact partition function  
$\mathcal{Z}(\beta) = 2^K (1 + (2^{\nL-\kL}-1)e^{-\beta})^{N_v}$: the syndrome on each node is independent, and there are $2^K$ configurations with each global syndrome. This is analytic as a function of temperature, so the equilibrium free energy does not detect any phase transition. Nevertheless, sharp phase transitions arising from a complex free energy landscape are witnessed by other probes, both in the dynamics (relaxation time) and the equilibrium behavior (the point-to-set correlation length)~\cite{Martinelli2004,Berger2005,montanari2006,montanari2006bethe,mezard2009information,Hong2025,ising}. 

The rugged free energy landscape in this model comes from the interplay between local constraints and geometry: random regular graphs are \textit{expander graphs}, which informally means that the boundary of a region scales proportionally to its bulk~\cite{expansion}. Sufficiently strong spectral expansion of the graph, together with good enough parameters on the local code, in turn implies \textit{code} expansion~\cite{sipser_spielman1996,guruswami2019essential,breuckmann2021balanced}: extensive energy barriers around ground states. 
Ref.~\cite{Placke_glass} proved that cLDPC codes with a subextensive number of redundancies and sufficiently strong code expansion have a spin glass phase at low temperatures. In that work and this one, the Gibbs state $p_\mathrm{G}(\vec{\sigma})$ is said to exhibit ``spin glass order'' if (i) it \textit{shatters} into exponentially many \textit{pure states}~\cite{van_enter1984stat_mech,georgii2011gibbs,mezard2009information,dembo_montanari2010review} (``typical'' valleys surrounded by free energy barriers), none of which carries more than an exponentially small fraction of the total weight, (ii) the configurational entropy density $\sconf(T)$, that is the Shannon entropy of the pure state decomposition~\cite{mezard2009information}, is strictly greater than the ground state entropy density for $T>0$ in the spin glass phase. 
Requiring property (ii), which we call \textit{incongruence}~\cite{incongruence}, means that spontaneous symmetry breaking of an exponentially large symmetry group $\mathbb{Z}_2^K$ does not alone qualify as glassiness: rather, the low-temperature landscape of an incongruently shattered system is dominated by exponentially many valleys containing  \textit{local} minima, not related by symmetries to those containing ground states. 

Our goal in this work is to characterize these valleys as a function of temperature and energy. With the appropriate choice of boundary conditions (on a tree) or initial conditions (for dynamics on a closed graph), we can put the system in one of its valleys, and it will remain in that valley as temperature is increased, up to a nonzero temperature (canonical instability). Alternatively, we can work in the \textit{microcanonical} ensemble, initializing the dynamics in a valley and finding the energy density at which it escapes that valley while staying at the same energy density (microcanonical instability). The theory is general, but for concreteness we will focus on the ``Tanner-Hamming [7,4,3] code'', whose local code has the parity check matrix~\cite{guruswami2019essential,tanner}

\begin{equation}\label{eq:hamming}
    H_{\rm L} = \begin{pmatrix} 1 & 1 & 1 & 0 & 1 & 0 & 0 \\ 0 & 1 & 1 & 1 & 0 & 1 & 0 \\ 1 & 1 & 0 & 1 & 0 & 0 & 1
    \end{pmatrix}.
\end{equation}
In the following, we use Ising variables, changing variables from bits $b=0,1$ to spins $\spin = 1 - 2b$. Notably, neither code expansion nor subextensive redundancies are proven for the Tanner-Hamming [7,4,3] code, leaving us without a formal proof of spin glass order in this model. 

\emph{Solution via tree recursion.---}
Since HGRRGs are locally tree-like, one way to study them is via recursion on rooted trees with different ensembles of boundary conditions (configurations on the leaves of the tree). This approach falls under the broad umbrella of \textit{cavity method}~\cite{Mezard1986cavity,mezard1987spin,mezard2001bethe,mezard2009information,mezard2015cavity}. 
Here we discuss the bulk energy, free energy, and entropy for a weighted ensemble of boundary conditions (BCs) parameterized by $(\alpha,\gamma)$. The boundary configuration $\vec{\spin}_\partial$ is weighted by a factor $(\mathcal{Z}_{\vec{\spin}_\partial}(\gamma\beta/\alpha))^\alpha$, where $\mathcal{Z}_{\vec{\spin}_{\partial}}(\gamma\beta/\alpha)$ is the partition function of a tree at inverse temperature $\gamma\beta/\alpha$, whose leaves are frozen to $\vec{\spin}_{\partial}$. In the one-step replica symmetry breaking (1RSB) formalism, $\gamma=\alpha \leq 1$ is a variational parameter known as the ``1RSB Parisi parameter''~\cite{mezard1987spin}, which is pinned to 1 in the dynamical 1RSB phase~\cite{franz2001ferromagnet,mezard2006reconstruction,krzakala2007gibbs,Krzakala2010,Zdeborova2010,mezard2009information}, but we will instead take $\gamma, \alpha$ to be separate tuning knobs which, at low enough bulk temperature $1/\beta$, adjust 
what set of valleys are being occupied by the system~\cite{alpha}. 
This family of BCs generalizes those considered in Ref.~\cite{Placke_glass}: ``codeword-polarized'' BCs ($\gamma = \infty,\alpha>0$) which force the system into the valleys containing ground states, and ``typical'' boundary conditions ($\alpha = \gamma = 1$)  which mimic the point-to-set construction on the closed graph~\cite{mezard2006reconstruction}. 

Consider the $\nL$ spins around a vertex. Designate one of these spins the ``output'' $\spin_0$, which is the one of these spins that is farther from the leaves of the tree, and the rest a $(b=\nL-1)$-component vector $\vec{\spin}$, such that the energy of this vertex for spin configuration $(\spin_0,\vec{\spin}) = (\spin_0,\spin_1,...,\spin_b)$ is $E_v(\spin_0,\vec{\spin})$. Likewise, let $(m_0,\vec{m})$ denote a $(b+1)$-component vector of magnetizations, where $m_i$ is the conditional magnetization at the root of a tree. 

Define
\begin{equation}\label{eq:z}
z(\vec{m}, \beta) = \sum_{\vec{\spin}}  \prod_{i=1}^b \left(\frac{1 + \spin_i m_i}{2}\right) \sum_{\spin_0} \exp(-\beta E_v(\spin_0, \vec{\spin})),
\end{equation}
which is the ``local partition function'' associated with joining together $b$ rooted trees with magnetizations $\vec{m}$. The conditional magnetization $m_0$ at the root of the resulting tree is then

\begin{align}\label{eq:recursion}
m_0=g(\vec{m},\beta)
&= \frac{\sum_{\spin_0,\vec{\spin}} \spin_0 e^{-\beta E_v(\spin_0,\vec{\spin})} \left(\prod_{i=1}^b \frac{1+\spin_i m_i}{2}\right)}{z(\vec{m},\beta)}.
\end{align}

Let $\Q{(r)}$ denote the distribution of the root magnetization for a given $\beta,\alpha,\gamma$, at depth $r$ from the leaves~\cite{fp}.  
$\Q{(r)}$ obeys the recursion relation
\begin{align}\label{eq:family}
&\Q{(r+1)}(m) \propto \int \delta(m - g(\vec{m},\beta)) z(\vec{\mu}, \gamma\beta/\alpha)^{\alpha} \prod_{i=1}^b d \Q{(r)}(m_i),
\end{align}
where $\vec{\mu}$ are the magnetizations obtained from the same boundary conditions as $\vec{m}$, but evaluated at inverse temperature $\gamma\beta/\alpha$, i.e. $\vec{\mu} = \vec{m}(\gamma\beta/\alpha)$.  

As derived in~\autoref{app:free-energy} of the Supplemental Information~\cite{supp-ref}, the bulk free energy density is
\begin{equation}\label{eq:free-main}
f(\beta,\alpha,\gamma) = f_{\rm int}(\beta,\alpha,\gamma) - \frac{\nL}{2} f_{\rm delta}(\beta,\alpha,\gamma)~,
\end{equation}
where $f_{\rm int}$ is the average change in the free energy due to joining together $\nL$ branches at a common interaction vertex, and $f_{\rm delta}$ is the average change in the free energy due to identifying two rooted trees on the root spin.

On the other hand, the bulk energy density $\varepsilon(\beta,\alpha,\gamma)$ is the expectation value of $E_v$ at a single node, weighted by the partition function from introducing that node. Note that the standard thermodynamic relation $\varepsilon(\beta) = \partial(\beta f)/\partial\beta$, which holds for a fixed valley, does not generically hold for fixed $(\alpha,\gamma)$, since the population of valleys selected by $(\alpha,\gamma)$ BCs changes with temperature.

If we do not condition on the boundary (or equivalently, consider the {full Boltzmann distribution of the system} on the closed graph, which has no redundancies), the {system's thermodynamic properties are those of the 
paramagnet, even when we are at low temperatures in the spin glass phase.} 
Within the 1RSB ansatz, the entropy density $s_{\rm para}$ of this paramagnet in general has two contributions: an ``intervalley'' (configurational) entropy density $\sconf$ from the multiplicity (if any) of typical valleys at that temperature, and an ``intravalley'' contribution from within each such valley. The entropy density with $\gamma=\alpha=1$ boundary conditions only receives the latter contribution, so the former, also called the \textit{complexity}~\cite{mezard2006reconstruction}, can thus be isolated as the difference~\cite{complexity}
\begin{equation}
\sconf(\beta) \equiv s_{\rm para}(\beta) - s(\beta, 1, 1)~,
\end{equation}
which is nonzero in the spin glass phase.

\emph{Low-temperature expansion.---}
Specializing to the [7,4,3] Tanner-Hamming code, we perform a low-temperature expansion for the bulk free energy density and energy density (per vertex) in the small parameter $x=\exp(-\beta)$~\cite{supp-ref}. Up to and including terms of order $x^3, x^{\gamma+1}, x^{(\gamma/\alpha) + 1}$, both quantities are functions of only $x$ and the defect density
\begin{equation}\label{eq:defect-density}
    \tilde{z} \equiv 2^\alpha x^\gamma (1 + 9\alpha x^{\gamma/\alpha})~.
\end{equation}
Here, a defect is a single violated node, which cannot be inserted or removed by local spin flips.  In terms of these two parameters,
\begin{align}
\beta f(x,\tilde{z}) &=  -\frac{7}{2} x^2 -28 x^3 - \frac{7}{2} \tilde{z} \left[\log(2x) + 9x + ...\right] \label{eq:f-defect} \\
\label{eq:e-defect}
\varepsilon(x, \tilde{z}) &= 7x^2 + 84x^3 + \frac{7}{2} \tilde{z}(1+9x+...)~.  
\end{align}
To leading order, the complexity is therefore~\cite{Placke_glass}
\begin{equation}\label{eq:s-conf-x}
\sconf(x) = \frac{1}{2}\log(2)+ 7x(1 - \log(2x)) + O(x^2)~.
\end{equation}
Crucially, the complexity increases as a function of temperature at low temperatures, signaling that the low temperature phase of this model indeed exhibits incongruent shattering.

\emph{Spin glass transitions.---} As temperature increases, there is an interplay of two effects: the energy density of states within a given valley increases due to thermal excitations; and for fixed finite $(\alpha,\gamma)$, more defects are forced in by the boundary conditions, thus selecting different valleys.  Eventually, the valleys induced by a given $(\alpha,\gamma)$ become unstable, and the system transitions to the paramagnetic phase at $T_c(\alpha,\gamma)$.  
The family of transitions parameterized by $(\alpha,\gamma)$ generalize the spin glass ($\alpha=\gamma=1$) transition \Tglass~and memory/codeword-polarized ($\gamma=\infty$) transition $T_{\rm mem}$, associated with the failure to recover information encoded in typical valleys and ground states, respectively~\cite{Placke_glass,lit}.

The transition is generically first-order, 
with a discontinuity in the energy density. Except in fine-tuned cases, there is a cusp in the bulk energy and free energy densities as the transition is approached from the low-temperature side, pointing to two different mechanisms for the instability: (1) a proliferation of thermal excitations, 
leading to an upwards cusp in $\varepsilon$, and (2) an entropic instability of forced-in defects to move towards the boundary.  As defects move towards the boundary, they can become more dilute in the bulk, resulting in a bulk energy density decrease.  
When this latter effect is dominant, $\varepsilon(\beta,\alpha,\gamma)$ has a \textit{downwards} cusp at $T_c$. 
Since $\gamma=\alpha=1$ selects the valleys with the typical energy density at that temperature---i.e., the energy density set by the global partition function, $\mathcal{Z}(\beta)$, which is paramagnetic---the standard spin glass transition at \Tglass~comes with no cusp in $\varepsilon$. Assuming that $s(\varepsilon,1,1)$ is a smooth function of the energy density, there is no cusp in $f$, either. 

\emph{High energy valleys.---} Now we   map out a larger region of the phase diagram that is not accessible under the $(\alpha, \gamma)$ BCs. We do so by tuning boundary conditions to force the system into valleys at nonzero energy density even at $T=0$, which remain stable up to a nonzero transition temperature. 

Note that with ``codeword-polarized'' boundary conditions, every root has the same $|m|$, and at each node, the signs on the inputs are chosen to be compatible with a codeword. We can generalize beyond this construction to allow a finite set of possible nodes, trading off between those that insert defects (raising the energy of the putative valley), and those that ensure stability.

To demonstrate this idea, {which is discussed further in~\autoref{app:fine-tuned} of the Supplemental Material}, consider building a tree with three types of nodes. Type 0 nodes force a defect: i.e. the incoming branches are 
inconsistent with any codeword, so $m_0=0$. Type 1 nodes have an energetically preferred root state (no defect), but some of the inputs are $m_i=0$, such that flipping the root to excite the node produces an \textit{entropy} increase, even at $T=0$.  To prevent the divergence of entropy that would destabilize the valley at nonzero temperature, we also introduce type 2 nodes, which have an energetically favorable root state and no $T=0$ entropy increase if the root spin is flipped.

Varying the composition of the three node types yields a family of valleys whose energy $\varepsilon(T)$ and critical temperature $T_c$ can be determined (semi-)analytically. These valleys provide bounds on stable regions of the phase diagram, shown in~\autoref{fig:e-together}. 

\begin{figure}[t]
\centering
\includegraphics[width=0.9\linewidth]{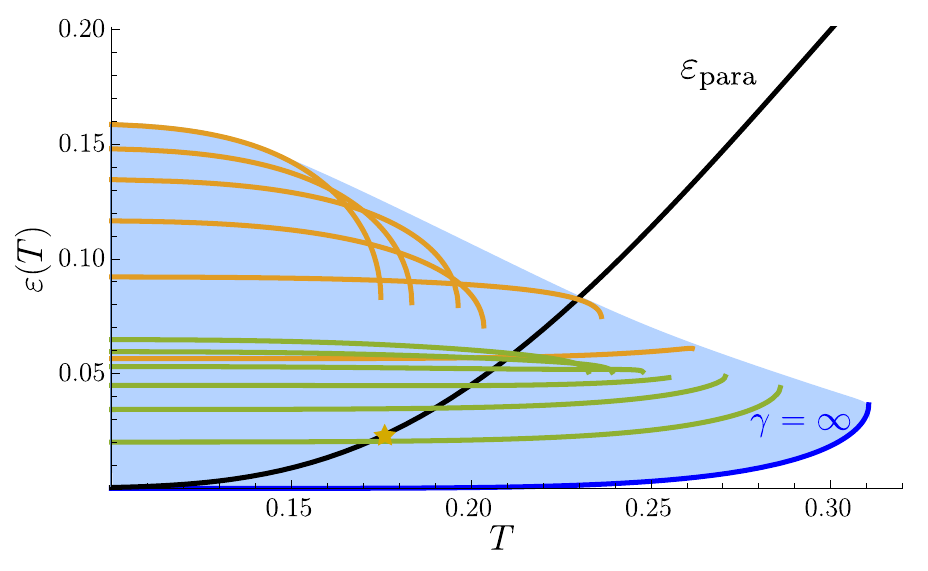}\label{fig:e-together}
\caption{Bulk energy density for valleys induced by various fine-tuned boundary conditions (orange and green curves). Also shown is the paramagnetic energy density $\ec{para}$ (black curve) and the energy density within codeword valleys ($\gamma=\infty$, blue curve). Star marks $(\Tglass, \varepsilon_{\rm G})$. Shading is conjectured region of thermally stable valleys that can be accessed with fine-tuned boundary conditions.
\label{fig:e-together}}
\end{figure}

\emph{Following valleys on closed graphs.---}
A natural question is whether the highest-energy valleys stabilized by fine-tuned boundary conditions on the tree are also accessible on closed graphs. The answer appears to be no: specially constructed states with high defect density on the closed graph appear to have low (order one) barriers and relax to lower $\varepsilon \lesssim 0.03$ when given local dynamics at low $T>0$. 

\begin{figure*}[t]
\includegraphics[width=\linewidth]{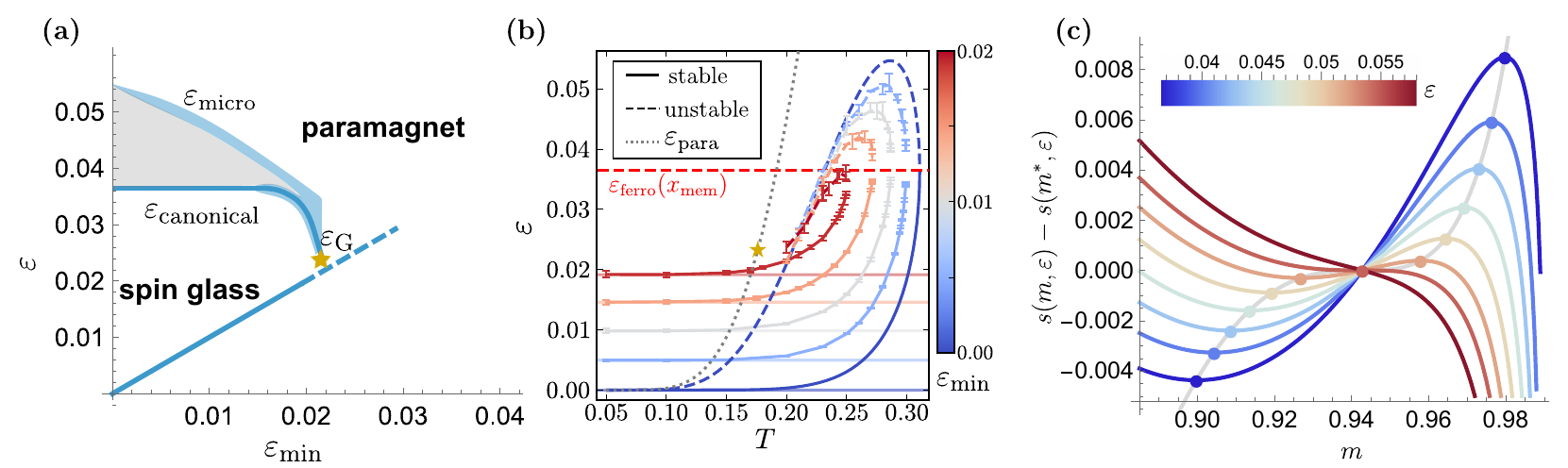}
\caption{(a) Phases of stability for valleys parameterized by $\ec{min}$, where $\ec{min}$ is the energy density at the bottom of the valley. Horizontal blue line marks the energy density at which a valley has a ``canonical instability''. Shaded gray region marks the regime where the valley is canonically unstable, but microcanonically stable due to an extensive entropy barrier to escape from the valley at that energy. Thick light blue curve is a rough estimate of the energy at which the entropic barrier vanishes; close to \ec{G}, the phase diagram is uncertain due to RSB within valleys. (b) Numerical evidence for $\ec{canonical}$. Solid (dashed) curves correspond to the energy at stable (unstable) fixed points associated with five sets of valleys: $\ec{0}=0,0.005,0.01,0.015,0.02$. $\ec{0}$ is the energy at which the solid curve crosses the paramagnetic energy curve (dotted gray). The five light horizontal lines indicate $\ec{min}(\ec{0})$, determined by an exact calculation~\cite{supp-ref}.The dashed horizontal line shows the analytically determined $\ec{canonical}(0)$. 
(c) Entropy barrier to escape valleys around ground states ($\ec{min}=0$). $m^*$ is the bulk magnetization at $\varepsilon=\ec{micro}(0)$. \label{fig:emin-escape}}
\end{figure*}

Nevertheless, there is a rich phase diagram of different valleys parameterized by their minimum energy density, $\ec{min}$. The phase diagram is sketched in~\autoref{fig:emin-escape}a. There are two ways that the system, initialized in a valley with a given $\ec{min}$, can escape the valley. We discuss these in turn. 

\emph{Canonical ensemble.---}
The first type of instability is via a ``canonical instability'' at $\ec{canonical}$: under local dynamics obeying detailed balance, a state initialized in that valley escapes to a much higher energy (leaving the valley) when the thermal expectation value of $\varepsilon$ exceeds $\ec{canonical}$. 

Local Metropolis dynamics initialized within a valley at inverse temperature $\beta$ samples states in that valley and produces a probability distribution of the energy $E$ that is proportional to $e^{-\beta F_{\rm valley}(E,T)}$. For $T<T_{\rm canonical}(\ec{min})$ and small enough $\ec{min}$, the within-valley free energy $F_{\rm valley}(E,T)$ has a local minimum within the valley that is surrounded by extensive free energy barriers; 
the canonical instability is when this local minimum vanishes, manifesting as an inflection point in $F_{\rm valley}(E,T)$ as a function of $E$. 

To detect this inflection point, we apply two complementary approaches. The first is an adaptive version of the local Metropolis algorithm on finite closed graphs, detailed in the End Matter. The second is a cavity method calculation inspired by Refs.~\cite{Krzakala2010,Zdeborova2010}. Briefly, consider the set of valleys that are typical at an inverse temperature $\beta_0$, where the corresponding energy density is $\varepsilon_0$. To adiabatically track this set of valleys as a function of temperature, we vary $\beta, \gamma$ together while fixing $\alpha=1, \gamma\beta = \beta_0$. Thus we trace out different curves $\varepsilon(\beta,\beta_0)$. In the limit $\beta\rightarrow \infty$, the recursion relation (\autoref{eq:family}) simplifies and we obtain an exact expression for the energy density at the bottom of the valley~\cite{supp-ref}. This simplified recursion relation admits nontrivial fixed points only for $T_0\leq 0.1718...\rightarrow \ec{min} \leq \ec{min}^* \approx 0.0195$, which provides an analytical lower bound on the spin glass temperature~\cite{Zdeborova2010}. 

The fact that $T_0$ falls slightly below our numerical estimate of $\Tglass=0.176(1)$ can be interpreted as follows~\cite{Zdeborova2010}. In the narrow interval $x_0 \in (x_0^*,x_{\rm G}]$, the system is unstable towards higher steps of replica symmetry breaking, so that this choice of BCs does not allow us to track an individual valley, which itself shatters into many subvalleys. This instability is fairly generic, having previously been observed in diluted $p$-spin models~\cite{montanari2003nature,montanari2004cooling_schedule,Zdeborova2010}.

On the other hand, valleys with $\ec{min}\leq \ec{min}^*$ are stable against RSB within the valley. We identify \ec{canonical} for these valleys by running population dynamics~\cite{Abou-Chacra_1973,mezard2001bethe,Mezard2003,mezard2006reconstruction} up to the temperature $T_c(\ec{min})$ at which the fixed point becomes trivial. Below $T_c$, we evaluate the bulk energy density $\varepsilon(\beta_0,\beta)$ at the stable nontrivial fixed point within the valley, as well as at the unstable fixed point where $F_{\rm valley}$ has its local maximum. $\varepsilon(\beta_0,\beta)$ and $\varepsilon_{\rm unstable}(\beta_0,\beta)$ have an upward and downward cusp, respectively, as $T\rightarrow T_c$, thus providing two-sided bounds on $\ec{canonical}$ (\autoref{fig:emin-escape}b). We find empirically that at small $\ec{min}$, $\ec{canonical}$ remains roughly equal to the value $7/192$ derived analytically for $\ec{min} = 0$, then decreases slightly as $\ec{min}^*$ is approached.

\emph{Microcanonical ensemble.---}
The second type of instability arises in the microcanonical ensemble. Consider the space of configurations $\{\vec{\sigma}\}$ with energy density $\varepsilon$. At sufficiently low $\varepsilon < \ec{micro}$, a state initialized within a given valley must pass through a sequence of rare configurations in order to escape the valley while staying at the same energy density, if it can escape at all. 

This phenomenon is best understood for the valleys containing ground states. The bulk energy density evaluated at the unstable fixed points of the recursion relation is nonmonotonic, reaching a maximum at $\ec{max}$. At energy density $\varepsilon<\varepsilon_{\rm{max}}$, there are two nontrivial solutions with magnetizations $m_1(\varepsilon) > m_2(\varepsilon)$ in the absence of a bulk field, and sweeping through an interval of magnetizations by applying a bulk field yields the entropy curve $s(m,\varepsilon)$ (\autoref{fig:emin-escape}c). $s(m,\varepsilon)$ has a local maximum at $m_1$, which we therefore identify with the population \textit{inside} the valley, and a local minimum at $m_2$, which we identify with saddle to escape the valley at that energy. At very low energies, $s(m_2,\varepsilon)<0$, signaling that although this unstable fixed point is realizable on a tree with a particular boundary, it is exponentially rare on a closed  random regular graph. We therefore tentatively identify $\ec{escape} \approx 0.0129$, where $s(m_2,\ec{escape})=0$, as the minimal energy at which a path to escape the valley exists. For $\ec{escape}\leq \varepsilon < \ec{max}$, paths to escape exist but are entropically suppressed: the microcanonical escape time is exponential in the entropic barrier $\Delta s(\varepsilon) = s(m_1,\varepsilon)-s(m_2,\varepsilon)$, which decreases as $\varepsilon$ increases until vanishing at $\ec{micro}=\ec{max}$.

This suggests a general procedure for identifying $\ec{micro}(\ec{min})$, using the data presented in~\autoref{fig:emin-escape}b: $\ec{micro}$ is simply the maximum energy density attained on the unstable branch, and is therefore strictly greater than  $\ec{canonical}$ for $\ec{min} \leq \ec{min}^*$. Thus, there is an intermediate energy regime wherein the valley is canonically unstable but microcanonically stable, as the only thermodynamically relevant escapes are to much higher energy density.
As $\ec{min}$ increases, the gap between $\ec{micro}$ and $\ec{canonical}$ decreases. For $\ec{min}>\ec{min}^*$, both $\ec{canonical}$ and $\ec{micro}$ become ill-defined within the 1RSB ansatz, leaving this region of the phase diagram uncertain. 

\emph{Discussion.---} The details of the energy landscape have direct consequences for error correction. Valleys corresponding to local minima with $\ec{min}>0$ can be used to enhance the effective code rate, at the expense of a lower threshold temperature for passive error correction. We also observe that up to high energy, the valleys around ground states are ``simple'': zero-temperature single-spin dynamics initialized within the valley succeeds in finding the ground state. From an error correction perspective, this means that one can successfully decode via the flip decoder, which seeks to lower the energy in every step~\cite{sipser_spielman1996}. Valleys around higher-energy local minima appear to have more ridges, suggesting that decoding instead be performed via low-temperature or the adaptive multicanonical dynamics detailed in the End Matter.

The numerical study of the Tanner-Hamming code opens up several avenues for future research. Most notably, the details of the landscape for $\ec{min}^* < \ec{min} < \ec{G}$ remain open. What is the nature of the transition at $T=T_\mathrm{G}$? Does the entropy barrier vanish at $\ec{G}$, or does it perhaps remain extensive? What is the highest energy density among the local minima (not necessarily typical) surrounded by extensive barriers in the energy landscape?  

While our numerics here have focused on locally tree-like graphs, where we can avail ourselves of the cavity method, the proof of spin glass order in Ref.~\cite{Placke_glass} applies to loopy graphs as well, and numerical simulations of the the Tanner-Hamming code on a hyperbolic lattice indicate that this model also has a spin glass and memory transition. A more detailed study of this phase diagram, using the flat histogram method, could shed light on possible qualitative differences between the landscape on locally tree-like vs. loopy graphs. To this end, developing a 1RSB ansatz for the corner transfer matrix renormalization group could be useful~\cite{Nishino1996,Ueda2007,Mosko2025,Wang2025}.

The simulation methods discussed in this work can also be used to study the \textit{quantum} LDPC codes, which are built from products of two cLDPC codes~\cite{tillich2009hgp,leverrier2015quantum_expander,breuckmann2021balanced,Breuckmann2021,leverrier2022quantum,panteleev2022asymptotically,dinur2023qldpc,Rakovszky2024} and support \textit{topological quantum spin glass order}~\cite{placke2024tqsg}.
The large system sizes in the quantum code families make precision numerics more challenging, but it would be interesting to find evidence of a similar thermally unstable, microcanonically stable regime as in~\autoref{fig:emin-escape}a. 

\emph{Acknowledgments.---} We thank Tibor Rakovszky, Vedika Khemani, and Nikolas Breuckmann for helpful discussions, collaboration on related work, and feedback on the manuscript. We especially thank Benedikt Placke for collaboration in the early stages of this project, including code and figures, and comments on the manuscript. We also acknowledge discussions with Federico Ricci-Tersenghi and David Gamarnik.  This work was supported in part by NSF QLCI grant OMA-2120757.  Numerical work was completed using computational resources managed and supported by Princeton Research Computing, a consortium of groups including the Princeton Institute for Computational Science and Engineering (PICSciE) and the Office of Information Technology's High Performance Computing Center and Visualization Laboratory at Princeton University. GMS additionally acknowledges the use of GPT-5.6 to format Algorithms 1 and 2 in the Supplemental Material and assist with coding in the late stages of the project.

\section*{End Matter}
This Appendix outlines the methods for direct simulations of the closed graph dynamics, which further support the phase diagram in~\autoref{fig:emin-escape}. By ``closed graphs'', we mean finite-size instances of HGRRGs, where $N_v=7778$, girth 6 is the largest accessible system size. A full description of the algorithms, along with details on the initialization within valleys and finite size effects, is provided in~\autoref{app:multi} of the Supplemental Material. 

\emph{Canonical instability.---} To identify the canonical instability, we use an adaptive local Metropolis algorithm, sometimes called the multicanonical method~\cite{Berg1991,Berg1992,Janke1998,Mau1999}, to efficiently sample states across a wide range of energies in the presence of extensive free energy barriers. We replace the Boltzmann factor $x=e^{-\beta}$ with the function $x(E)$, adjusted so that histogram of energies follow a particular functional form. We also impose a cutoff on the space of allowed configurations to prevent the system from escaping the valley~\cite{supp-ref}. Then, a spin flip move which would change the energy from $E$ to $E+\Delta E$ (where $|\Delta E| \leq 2$) is rejected if it takes the state outside the space of allowed configurations, and otherwise accepted with probability
\begin{equation}\label{eq:P-accept}
    P_{\rm accept} =\begin{cases} 1 & \Delta E \leq 0~, \\
    \prod_{i=1}^{\Delta E} x(E + i) & \Delta E > 0~.
    \end{cases}
\end{equation} 
If we choose $x(E)$ adaptively so that energies within a chosen interval are sampled with equal probability (i.e., the energy histogram is flat on this interval), then $x(E) = \exp[-S'(E)]$, and thus the inflection point in the free energy manifests as a local maximum in $x(E)$. 
We can then read off
\begin{subequations}
\begin{align}
    \ec{canonical} &= \mathrm{argmax}[x(E)]/N_v \\
    T_{\rm{canonical}} &= -1/\log(x(\ec{canonical} N_v))~.
\end{align}
\end{subequations}
With increasing system size, $\ec{canonical}$ and $T_{\rm canonical}$ appear to be converging towards the values obtained via the cavity method in the main text. In particular, for small $\ec{min}$, $\ec{canonical}$ inferred from the largest accessible system size $N_v=7778$ is in good agreement with $\ec{ferro}(x_{\rm mem})$, 
where $T_{\rm mem}$ is the memory/codeword-polarized transition temperature.

\emph{Microcanonical instability.---} To probe the microcanonical instability, we restrict the space of allowed configurations to a narrow window centered on energy $E$. We organize the high-dimensional configuration space within this window by a single additional coordinate, $\vec{\sigma} \rightarrow M(\vec{\sigma})$, defined as the average correlation between the configuration $\vec{\spin}$ and the states at the bottom of the chosen valley. For a given $E$, the density of states $n(E,M) = e^{S(E,M)}$ has a local maximum at $M^*(E)$, corresponding to the set of typical states within the valley at that energy, and a local minimum at $M_{\rm barrier}(E) < M^*(E)$.  $\Delta S(E) \equiv S(E,M^*(E)) - S(E,M_{\rm barrier}(E))$ is a lower bound on the true entropic barrier~\cite{supp-ref}, so identifying the energy density at which $\Delta S$ vanishes furnishes a lower bound on the microcanonical phase boundary.

To measure $\Delta S$, we again employ adaptive Metropolis techniques, this time with transition probabilities set by $M$ instead of $E$. The finite size effects for the valleys containing ground states are stronger than in the canonical ensemble, and it is numerically challenging to probe $\ec{micro}$ for higher-energy valleys. However, where we have been able to obtain data, it is broadly consistent with or converging towards the tree calculations of the main text.

%% file: main.bbl
%

%% file: supp-clean.tex
\title{Supplemental Information to: ``Stable valleys in the glassy landscape of a low-density parity-check (LDPC) code''}
\maketitle
\onecolumngrid

The Supplemental Material is organized as follows.

\begin{itemize}
    \item \autoref{app:prelim} (re-)introduces the formalism and notation for tree calculations and outlines the population dynamics method.
    \item \autoref{app:bulk} elaborates on the calculation of bulk observables on the Bethe lattice with $(\gamma,\alpha)$ boundary conditions.
    \item \autoref{app:prior} reviews prior results on the Tanner-Hamming model from Ref.~\cite{Placke_glass} and compares the model to the cavity method literature on XOR-SAT/diluted $p$-spin models/Gallager codes.
    
    \item \autoref{app:low-T} derives the low-temperature series expansion for $(\gamma,\alpha)$ boundary conditions and complements it with numerical simulations.
    \item \autoref{app:fine-tuned} details the construction of fine-tuned boundary conditions which induce high-energy valleys (\autoref{fig:e-together} of the main text).
    \item The remaining sections discuss the two complementary approaches used to obtain the phase diagram of canonical and microcanonical instabilities (\autoref{fig:emin-escape} of the main text):
    \begin{itemize}
        \item \autoref{app:alpha1} expands on the $\alpha=1, \gamma\beta=\beta_0$ cavity method.
\item \autoref{app:multi} describes the adaptive histogram method for Metropolis dynamics on closed graphs. We also present supporting numerics on the entropy barriers.
\end{itemize}
\end{itemize}

\section{Preliminaries}\label{app:prelim}
\subsection{Notation}
As in the main text, we consider a local code involving $\nL=b+1$ spins and fix a labeling of spins around each node. On the tree, it is convenient to consider one of the legs an ``output'' $\spin_0$ and the rest ``inputs'' $\vec{\spin}$, such that each node is assigned an energy $E_v(\spin_0,\vec{\spin}) \in \{0, 1\}$. 

Consider a rooted tree whose leaves are frozen to the configuration $\vec{\spin}_\partial$. The magnetization at the root, conditioned on the leaves, is 
\begin{equation}
m_{\vec{\spin}_\partial} = \frac{\mathcal{Z}_{+1,\vec{\spin}_\partial}(\beta) - \mathcal{Z}_{-1,\vec{\spin}_\partial}(\beta)}{\mathcal{Z}_{\vec{\spin}_\partial}(\beta)}
\end{equation}
where $\mathcal{Z}_{\spin,\vec{\spin}_\partial}$ is the partition function on the tree with the root and leaves frozen to $\spin, \vec{\spin}_{\partial}$ respectively, and $\mathcal{Z}_{\vec{\spin}_\partial}$ is the partition function leaving the root free.

Now let $(m_0,\vec{m})$ denote a $(b+1)$-component vector of magnetizations, where $m_i$ is the conditional magnetization at the root of the $i$th tree before introducing the interaction on node $v$. If we join together branches $1,2,\dots,b$, then the conditional magnetization at the new root $0$ is $g(\vec{m},\beta)$ (see \autoref{eq:recursion} and~\autoref{fig:recursion}a).

For ease of notation, when $m_i=m$ for all $i=1,...,b$ (identical ``inputs''), we will write $g(m,\beta)$ and $z(m,\beta)$ in place of $z(\vec{m},\beta)$ and $g(\vec{m},\beta)$. As long as the local code does not have any single-body checks, the linearity of the code implies that there are an equal number of local codewords with $\sigma_0=1$ and $\sigma_0=-1$, which in turn implies that $m=0$ is a fixed point of the equation $m = g(m,\beta)$. 

If the local code encodes $\kL$ logical bits into $\nL$ physical bits (that is, $\nL$ spins are involved in $\nL-\kL$ independent constraints), then (adopting the notation $Z(\beta) = z(0,\beta)$) we obtain
\begin{equation}
Z(\beta) = 2^{\kL-\nL+1} (1 + (2^{\nL-\kL}-1)x).
\end{equation}
In the absence of redundancies, the partition function on a closed random regular graph with $N_v$ vertices is 
\begin{equation}\label{eq:Zpara}
\mathcal{Z}(\beta) = 2^{N_v(\nL/2-1)}Z(\beta)^{N_v} = 2^K\left(1 + (2^{\nL-\kL}-1)x\right)^{N_v}.
\end{equation}

This ``global'' Gibbs state is paramagnetic: it has short-ranged correlations and no spontaneous symmetry breaking at any temperature. From~\autoref{eq:Zpara} we can read off the paramagnetic free energy, energy, and entropy density per node:

\begin{subequations}\label{eq:efs-para}
\begin{align}
\beta f_{\rm para}(x) &= (1-\nL/2)\log(2) - \log(Z(\beta)) =-R \log(2) -\log\left[1 + (2^{\nL-\kL}-1)x)\right], \label{eq:f-para} \\
\varepsilon_{\rm para}(x) &= \frac{\partial}{\partial \beta} (\beta f_{\rm para}) = -\frac{\partial}{\partial \beta} \log Z(\beta) =\frac{x (2^{\nL-\kL}-1)}{1 + (2^{\nL-\kL} - 1) x}, \label{eq:e-para} \\
\quad s_{\rm para}(x) &= \beta(\varepsilon_{\rm para}(x) - f_{\rm para}(x)),
\end{align}
\end{subequations}
where $R=K/N_v$ is the log base 2 of the ground state degeneracy per node.\footnote{In error correction, the \textit{rate} of a code is commonly defined as the ratio of the number of logical bits [for us, $K=N_v(\kL-\nL/2)$] to the number physical bits (for us, $N_v \nL/2$), which evaluates to $2\kL/\nL - 1$. In this work, we define all densities as the density per node rather than density per spin, so the code rate per node, that is the entropy density at $T=0$, is simply $R=\kL-\nL/2$.}

Throughout, we will use lowercase letters $f,\varepsilon,s,m$ to denote intensive quantities, and uppercase letters $F,E,S,M$ to denote extensive quantities (e.g. on finite graphs). When discussing these quantities as a function of temperature, we will switch freely between the argument $T$, $\beta=1/T$, and $x=e^{-\beta}$ depending on context.

\subsection{Specializing to the Tanner-Hamming code}
If the local code is the [7,4,3] Hamming code, then
\begin{align}
z(\vec{m},\beta) &= \frac{1}{4}\left[1 + 7 x + (1-x) (m_2 m_3 m_4 m_6 + m_1 m_5 m_2 m_3 + m_1 m_5 m_4 m_6)\right] \label{eq:z} \\
g(\vec{m},\beta) &= \frac{(m_1 m_2 m_4 + m_3 m_4 m_5 + m_1 m_3 m_6 + m_2 m_5 m_6) (1 - 
   x)}{4 z(\vec{m},\beta)}, \label{eq:f}
\end{align}
where $x=e^{-\beta}$. 

With identical inputs,~\autoref{eq:z} simplifies to
\begin{equation}\label{eq:z-ferro}
    z(m,\beta)=\frac{1}{4}(1+7x + 3m^4(1-x)),
\end{equation}
while~\autoref{eq:f} simplifies to
\begin{equation}\label{eq:fm-ferro}
g(m,\beta) = \frac{4 m^3 (1 -x)}{1 + 7x + 3 m^4 (1-x)}.
\end{equation}

The local code has $\nL=7,\kL=4$, so the code rate per node evaluates to $R=1/2$.

For completeness, we state here the 16 codewords of $H_{\rm L}$:
\begin{enumerate}
    \item 2 codewords with $|\sum_i \spin_i| = \nL$: the all up codeword and its $\mathbb{Z}_2$ partner, the all down state
    \item 7 codewords with $\sum_i \spin_i = 1$: $\begin{pmatrix} 1 & 1 & 1 & -1 & 1 & -1 & -1 \end{pmatrix}$ and cyclic permutations
    \item 7 codewords with $\sum_i \spin_i = -1$, obtained by flipping all spins in the codewords of set (2).
\end{enumerate}

\subsection{Ensembles of boundary conditions}
Recall that we defined $\Q{(r)}$ as the probability distribution of root magnetizations in a depth $r$ rooted tree, where the distribution is taken over an ensemble of boundary conditions weighted by $\mathcal{Z}_{\vec{\spin}_\partial}(\gamma\beta/\alpha)^\alpha$. Provided that the local code has no single-body checks, $\Q{}$ has the $\mathbbm{Z}_2$ symmetry $\Q{}(m)=\Q{}(-m)$.

In the main text, we gave the distributional recursion relation (\autoref{eq:family}) up to proportionality; here we make the proportionality factor explicit:

\begin{align}\label{eq:family-normalized}
\Q{(r+1)}(m) = \frac{1}{\mathcal{Z}_{\rm iter}(\beta,\alpha,\gamma)}
\int \delta(m - g(\vec{m},\beta)) z(\vec{\mu}, \gamma\beta/\alpha)^{\alpha} \prod_{i=1}^b d \Q{(r)}(m_i)
\end{align}
where
\begin{equation}\label{eq:Z-iter}
\mathcal{Z}_{\rm iter}(\beta,\alpha,\gamma) \equiv \int [z(\vec{\mu}, \gamma\beta/\alpha)]^\alpha \prod_{i=1}^b d\Q{(r)}(m_i).
\end{equation}

It will sometimes be useful to consider instead the distribution $\tQ{s}{}$, defined via the relation
\begin{equation}
\tQ{\spin}{}(m) = (1 + \mu \spin) \Q{}(m) \Leftrightarrow \Q{} = \frac{1}{2} \left(\tQ{1}{} + \tQ{-1}{}\right),
\end{equation}
where $\mu(x) = m(x^{\gamma/\alpha})$. This is the distribution of conditional root magnetizations across the following ensemble: freeze the root to $\spin$ and sample configurations on the leaves, $\vec{\spin}_\partial$, with probability proportional to $\mathcal{Z}_{\spin,\vec{\spin}_\partial}(\gamma\beta/\alpha)^\alpha$, where $\mathcal{Z}_{\spin,\vec{\spin}_\partial}$ is the partition function on the tree with the root and leaves frozen to $\spin, \vec{\spin}_{\partial}$ respectively.

This alternative formulation is especially useful for $\alpha = 1$. Since $\Q{}(m)=\Q{}(-m)$, it follows that $\mathcal{Z}_{\rm iter}(\beta,1,\gamma) = Z(\gamma \beta)$, and the recursion relation for $\tQ{\spin_0}{(r)}$ becomes
\begin{equation}\label{eq:Qtilde}
\tilde{Q}_{\spin_0,\beta,1,\gamma}^{(r+1)}(m) = \frac{1}{2^{b-1} Z(\gamma\beta)} \sum_{\vec{\spin}} \int \delta(m - g(\vec{m},\beta)) e^{-\gamma \beta E_v(\spin_0, \vec{\spin})} \prod_{i=1}^b d\tilde{Q}_{\spin_i,\beta,1,\gamma}^{(r)}(m_i).
\end{equation}
From this perspective, at $\alpha=1$, we are ``broadcasting'' a message from root to leaves at inverse temperature $\gamma\beta$, and ``decoding'' it at inverse temperature $\beta$. As argued in Refs.~\cite{Krzakala2010,Zdeborova2010}, fixing $\gamma \beta=\beta_0$ and varying $\gamma$ is a way to vary the temperature \textit{within} a certain valley (sampled from the equilibrium measure inverse temperature $\beta_0$). We exploited this fact to obtain the phase diagram in~\autoref{fig:emin-escape}, and elaborate upon the attendant calculations in~\autoref{app:alpha1} of this Supplement. 

\subsection{Population dynamics}\label{app:pop-dynamics}
To perform numerics on trees, we use population dynamics~\cite{Abou-Chacra_1973,mezard2001bethe,Mezard2003,mezard2006reconstruction}, a technique for sampling the distributional recursion relations that arise in the cavity method.

For general $\gamma, \alpha$, we simulate~\autoref{eq:family} as follows. Initialize a population of $2M$ triples, $(m,\mu, w)$, where $m$ is a conditional magnetization, $\mu$ is the magnetization obtained from the same boundary conditions but evaluated at inverse temperature $\gamma \beta/\alpha$, and $w$ is the weight of this sample in the population. As the initial condition, take $m=\mu=1$ in $M$ samples, and $m=\mu=-1$ in the remaining $M$ samples, and assign each sample equal weight. Then, for each $r=0,...,r_{\rm max}$, and for $j=1,\dots,2M$,
\begin{enumerate}
\item Draw $(m_i,\mu_i)$, $i=1,\dots,b$, independently from the weighted population $\Q{(r)}$, with probability proportional to the weight $w_i$. Let $\vec{m}=(m_1,\dots,m_b)$ and $\vec{\mu}=(\mu_1,\dots,\mu_b)$.
\item Set the $j$th element of $\Q{(r+1)}$ equal to $(m,\mu,w)$ where
\begin{align}
m = g(\vec{m},\beta), \quad \mu = g(\vec{\mu}, \gamma \beta/\alpha), \quad w=z(\vec{\mu},\gamma\beta/\alpha)^\alpha \prod_{i=1}^b w_i.
\end{align}
\end{enumerate}

There are two simplifying cases. First, if $\gamma=\alpha$, then $m=\mu$ and we only need to store tuples $(m,w)$. Second, if $\alpha=1$, then we can follow the recursion using two \textit{unweighted} populations. The trick is to consider $\tilde{Q}$ instead of $Q$, so that factors of $z(\vec{\mu},\beta)$ cancel out in the recursion relation. We simulate~\autoref{eq:Qtilde} as follows. Initialize two populations of size $M$, $\tilde{Q}_{1,\beta,1,\gamma}$ and $\tilde{Q}_{-1,\beta,1,\gamma}$. The initial condition is $m=\mu=\sigma_0$: decoding is perfect in a depth 0 tree. Then, for $\spin_0=\pm 1$ and $j=1,...,M$,

\begin{enumerate}
\item Sample the spin configuration $\vec{\spin} = \spin_1,\dots,\spin_b$ with Boltzmann weight $e^{-\gamma \beta E_v(\spin_0,\vec{\spin})}$.
\item For each $\spin_i$, $i=1,\dots b$, independently sample $m_i$ from population $\tilde{Q}^{(r)}_{\spin_i,\beta,1,\gamma}$.
\item Set the $j$th element of the population $\tilde{Q}^{(r+1)}_{\spin_0,\beta,1,\gamma}$ equal to $(m,\mu)$ where $m=g(\vec{m},\beta), \mu = g(\vec{\mu},\gamma\beta)$.
\end{enumerate}

To analyze properties of stable fixed points, the process is iterated up to a predetermined $r_{\rm max}$, chosen large enough so that $Q$ or $\tilde{Q}$ has reached an approximate steady state, as indicated by a plateau in $\langle |m|\rangle$. The population size is $M=10^5$, and expectations values are taken over $\geq 20$ independent runs. As described in the next section, sampling of bulk observables requires a final step that translates from root to bulk properties. When working with populations of $\tilde{Q}$, we perform this final sampling step on the aggregated population $Q = \tilde{Q}_1 \cup \tilde{Q}_{-1}$. 

The analysis of unstable fixed points at $\alpha=1$ (dashed curves in~\autoref{fig:emin-escape}b) calls for a few modifications of the above procedure. First, rather than initializing $m=\mu=\pm 1$, which would flow toward the stable fixed point, we sweep through a range of $m$ and identify the initial conditions that are sufficiently close to an unstable fixed point. Here, by ``sufficiently close'', we mean that some fraction of runs flow towards an approximate plateau before being repelled towards the nontrivial stable fixed point, and the remaining fraction flow towards an approximate plateau before being repelled towards the paramagnetic fixed point. We estimate $\ec{unstable}$ as the approximate dividing line in energy between these two groups of runs. We further exploit the fact that, once the integrals in the next section are rewritten as integrals over conditional distributions $\tilde{Q}$, the dependence on $\mu$ drops out. This way of sampling observables incurs larger statistical errors than the method of the next section, but it conveniently avoids the problem of how to initialize $\mu$. Finally, since the temporary plateau is short-lived, we also modify the definition of one iteration to consist of replacing one randomly selected element of the current population with a newly generated element, rather than updating the entire population.

\section{Calculation of bulk observables on the Bethe lattice}\label{app:bulk}
\begin{figure}[tb]
\centering
\includegraphics[width=0.7\linewidth]{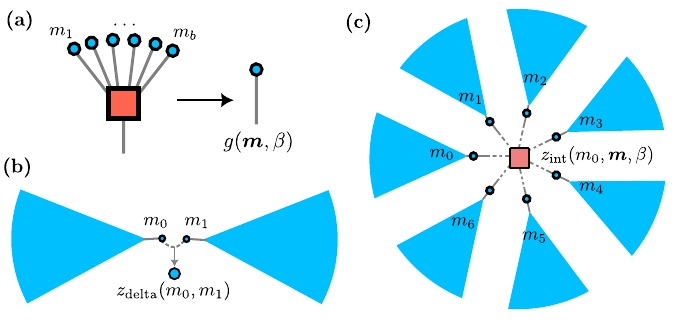}
\caption{Methods of recursion on trees. (a) Joining together $b$ branches with conditional magnetizations $\vec{m}$ $(m_1,...,m_b)$ yields a new rooted trees with conditional magnetization $g(\vec{m},\beta)$. (b), (c). Two ways to make an infinite Bethe lattice out of infinite rooted trees (blue wedges). In (b), we identify the root spins of two rooted infinite trees (blue wedges) which yields an infinite Bethe lattice with one fewer spin. The joining step has a local partition factor of $z_{\rm delta}(m_0,m_1)$ and a shift $Y_{\rm delta}(m_0,m_1)$ in the chosen observable $Y$ (e.g., the free energy density). In (c), we join $b+1$ branches together at a common interaction vertex, yielding an infinite Bethe lattice with one extra vertex node. If the branches have root magnetizations $m_0,\vec{m}$, then the shift in observable $Y$ is $Y_{\rm int}(m_0,\vec{m},\beta)$ and the local partition factor is $z_{\rm delta}(m_0,\vec{m},\beta)$.\label{fig:recursion}}
\end{figure}
\subsection{General method}
Our recursive methods for studying trees with a given (ensemble of) boundary condition allow us to determine, analytically or numerically, the fixed point distribution of conditional magnetization at the \textit{root} of the tree. To determine the \textit{bulk} density of a thermodynamic quantity which, for fixed boundary conditions, is expressed as a function of the partition function, we consider two equivalent ways of constructing a Bethe lattice from $2(b+1)$ infinite rooted trees:

\begin{enumerate}
\item Pair up the trees and join them at the root spins, making $b+1$ Bethe lattices with $b+1$ fewer spins than we started with. Pairing trees with magnetizations $m_0,m_1$ at the roots incurs a shift $Y_{\rm delta}(m_0,m_1)$ in the quantity of interest (\autoref{fig:recursion}b). 
\item Make two groups of $b+1$ branches and join the branches in each group at a single node, making 2 Bethe lattices with 2 more nodes than we started with. Joining branches with magnetizations $(m_0,\vec{m})$ at their roots incurs a shift of $Y_{\rm int}(m_0,\vec{m},\beta)$ (\autoref{fig:recursion}c).
\end{enumerate}
Therefore,
\begin{equation}\label{eq:Y-bulk}
Y_{\rm bulk} = Y_{\rm int} - \frac{b+1}{2} Y_{\rm delta}.
\end{equation}
For the ensemble of boundary conditions parameterized by $(\alpha,\gamma)$, the average shift $Y_{\rm delta}$ is
\begin{align}\label{eq:Y-delta}
    Y_{\rm delta}(\beta,\alpha,\gamma) &= \frac{\int ((1+\mu_0\mu_1)/2)^{\alpha} Y_{\rm delta}(m_0,m_1) d\Q{}(m_0) d\Q{}(m_1)}{\int ((1+\mu_0 \mu_1)/2)^{\alpha} d\Q{}(m_0) d\Q{}(m_1)} \notag \\
&\equiv \frac{\int z_{\rm delta}(\mu_0, \mu_1)^{\alpha} Y_{\rm delta}(m_0, m_1) d\Q{}(m_0) d\Q{}(m_1)}{\mathcal{Z}_{\rm delta}(\beta, \alpha,\gamma)}.
\end{align}
where $z_{\rm delta}(m_0,m_1) = (1 + m_0 m_1)/2$. 

Further defining  
\begin{equation}\label{eq:z-int}
z_{\rm int}(m_0, \vec{m}, \beta) = \sum_{\spin_0, \vec{\spin}} e^{-\beta E_v(\spin_0, \vec{\spin})} \prod_{i=0}^b \left(\frac{1 + \spin_i m_i}{2}\right) = z(\vec{m}, \beta) z_{\rm delta}(m_0, f(\vec{m},\beta))
\end{equation}
and
\begin{equation}\label{eq:Z-int}
\mathcal{Z}_{\rm int}(\beta,\alpha,\gamma) =  \mathcal{Z}_{\rm iter}(\beta,\alpha,\gamma)\mathcal{Z}_{\rm delta}(\beta,\alpha,\gamma),
\end{equation}
the average shift $Y_{\rm int}$ evaluates to
\begin{equation}\label{eq:Y-int}
Y_{\rm int}(\beta,\alpha,\gamma) = \frac{\int z_{\rm int}(\mu_0, \vec{\mu}, \gamma\beta/\alpha)^\alpha Y_{\rm int}(m_0,\vec{m},\beta) \prod_{i=0}^b d\Q{}(m_i)}{\mathcal{Z}_{\rm int}(\beta,\alpha,\gamma)}.
\end{equation}

\subsection{Free energy density}\label{app:free-energy}
The bulk free energy density is obtained by substituting $Y=\beta f$, where
\begin{equation}\label{eq:beta-f}
\beta f_{\rm delta}(m_0,m_1) = -\log(z_{\rm delta}(m_0,m_1)), \qquad \beta f_{\rm int}(m_0,\vec{m},\beta) = -\log(z_{\rm int}(m_0,\vec{m},\beta)).
\end{equation}

\autoref{eq:Y-bulk} together with~\autoref{eq:beta-f} may bring to mind the expression for the Bethe free energy on a factor graph~\cite{mezard2009information}. Here, $z_{\rm int}$ and $z_{\rm delta}$ correspond to local partition functions on factors and variables, respectively, which are expressed in terms of ``messages'' (i.e., conditional magnetizations) received from neighboring nodes. When the messages satisfy the so-called belief propagation (BP) equations, the Bethe free energy has a stationary point. When the underlying graph is a tree, the fixed point messages are uniquely determined by the configuration on the leaves, and the Bethe free energy at its stationary point is the exact free energy for that boundary condition~\cite{Yedidia2005}. 

However, as we have stressed,  different boundary configurations can lead to different fixed point messages. For an ensemble of boundary conditions, the message-passing equations become a distributional recursion relation [\autoref{eq:family-normalized}], and we must average over the resulting fixed point distribution of messages $\Q{}$ to obtain the ensemble-averaged bulk free energy density. \autoref{eq:Y-delta} and~\autoref{eq:Y-int} tell us how to take this average.\footnote{See Refs.~\cite{mezard2001bethe,mezard2006reconstruction} for a similar expression in the case $\gamma=\alpha=1$, applied to the Bethe lattice Ising/Potts model, and Ref.~\cite{Zdeborova2010} for the generalization to $\alpha=1$, arbitrary $\gamma$, applied to the diluted $p$-spin model.} The choice $\alpha=\gamma=1$ is well-motivated because the fixed point distribution of messages then yields a stationary point of the properly averaged free energy density ~\cite{mezard2006reconstruction}. Namely, from~\autoref{eq:Y-int},~\autoref{eq:Y-delta}, and~\autoref{eq:beta-f}, one can show that, letting $f(\lambda) = (1-\lambda) f(Q^*_{\beta,1,1}) + \lambda f(Q)$ for any distribution $Q$ and $\lambda \in [0,1]$,
\begin{equation}
\frac{df}{d\lambda}{\Large|}_{\lambda=0} = 0.
\end{equation}

\subsection{Configurational entropy density}
In the main text, we expressed the configurational entropy as a function of temperature, obtained by setting $\gamma=\alpha=1$ to select typical valleys. However, in the study of 1RSB, one is often interested in the configurational entropy as a function of free energy, commonly denoted $\Sigma(f)$ and defined as the logarithm of the density of pure states (valleys) at a given internal (intravalley) free energy density~\cite{mezard2009information}. This is obtained by viewing $\gamma=\alpha=y$ as a variational parameter (the Parisi parameter) conjugate to $f$. That is, following the presentation in Ref.~\cite{Zdeborova2010},
\begin{equation}\label{eq:conf-f}
    -\beta y \Phi(\beta,y) = -\beta y f(\beta,y) + \Sigma(f),
\end{equation}
where
\begin{equation}\label{eq:Phi}
    \beta y \Phi(\beta,y) = -\log \mathcal{Z}_{\rm int}(\beta,y,y) + \frac{b+1}{2} \log \mathcal{Z}_{\rm delta}(\beta, y, y).
\end{equation}
Here $y \Phi(\beta,y)$ is the total free energy, and per usual, the value of $y$ in equilibrium at inverse temperature $\beta$ is set by minimizing $y \Phi$ with respect to $y$. Hence 
we define
\begin{equation}
    f(\beta, y) = \frac{\partial [y \Phi(\beta, y)]}{\partial y},
\end{equation}
which, upon comparison to \autoref{eq:Y-delta} - \autoref{eq:beta-f}, indeed recovers our earlier expression for $f(\beta, \alpha,\gamma)$ [\autoref{eq:free-main}] along the line $\alpha=\gamma$. In the minimization of $y\Phi$, we technically must restrict to $y$ such that $\Sigma(f)$ is non-negative; in the static 1RSB phase, this constraint leads to $y^*<1$, whereas in the dynamical 1RSB phase, we maintain $y^*=1$. As discussed in~\autoref{app:prior} of this Supplement, our models do not exhibit static RSB, so we can safely set $y^*=1$, recovering $\Sigma(f(\beta,1)) = s_{\rm conf}(\beta)$: the configurational entropy density at some temperature is just log the density of states at the free energy density which is typical at that temperature. 
More generally, we can use~\autoref{eq:conf-f} and~\autoref{eq:Phi} to together define the complexity $\sconf(\beta,\alpha=\gamma)$ of valleys as a function of temperature for a given $\gamma=\alpha$, under the usual assumption that the valleys selected by $(\beta,\alpha,\gamma)$ BCs are sufficiently concentrated in free energy.
The limit $y\rightarrow\infty$ offers a simple sanity check. For $T<\Tdyn$, $f(\beta,\infty,\infty)$ is the internal free energy of the valleys containing codewords, and substituting \autoref{eq:ferro-fp} below into \autoref{eq:beta-f} and \autoref{eq:Phi} yields $\sconf(\beta,\infty)=R\log2$: the configurational entropy of valleys containing codewords is simply log2 times the bare code rate.

\subsection{Energy density}\label{app:energy-density}
The energy density can be evaluated in a similar way to the free energy density, defining $Y=\partial (\beta f)/\partial \beta$:
\begin{equation}\label{eq:e-two-parts}
    \varepsilon_{\rm delta}(m_0,m_1) = -\frac{\partial \log(z_{\rm delta}(m_0(\beta),m_1(\beta)))}{\partial \beta}, \qquad \varepsilon_{\rm int}(m_0,\vec{m}) = \frac{\partial \log(z_{\rm int}(m_0(\beta), \vec{m}(\beta), \beta))}{\partial \beta}
\end{equation}
Note that $m_i$ is treated as a function of $\beta$, so the derivative acts nontrivially on each argument of $z_{\rm delta}, z_{\rm int}$.

In the version of population dynamics outlined in~\autoref{app:pop-dynamics}, we only keep track of the root magnetizations, not their $\beta$ dependence. While we can recover the $\beta$ dependence by also tracking the difference $\varepsilon_i$ between the thermal expectation value of the energy with root +1 vs. root $-1$, there is fortunately a simpler way: defining
\begin{equation}\label{eq:Ev-ga}
\langle E_v (m_0, \vec{m}) \rangle \equiv -\frac{\partial}{\partial \beta} \log(z_{\rm int}(m_0, \vec{m}, \beta)) = \frac{\sum_{\spin_0, \vec{\spin}}e^{-\beta E_v(\spin_0,\vec{\spin})} E_v(\spin_0, \vec{\spin}) \prod_{i=0}^b (1 + m_i \spin_i)}{\sum_{\spin_0, \vec{\spin}}e^{-\beta E_v(\spin_0,\vec{\spin})} \prod_{i=0}^b (1 + m_i \spin_i)},
\end{equation}
where now the derivative acts only on the explicit argument $\beta$, some algebraic manipulation yields
\begin{equation}\label{eq:e-gamma-alpha}
\varepsilon(\beta, \alpha,\gamma) = \frac{\int \langle E_v(m_0, \vec{m}) \rangle z_{\rm int}(\mu_0,\vec{\mu},\gamma\beta/\alpha)^{\alpha} \prod_{i=0}^b d\Q{}(m_i)}{\mathcal{Z}_{\rm int}(\gamma,\beta,\alpha)}.
\end{equation}
As indicated by the second equality in~\autoref{eq:Ev-ga}, $\langle E_v(m_0,\vec{m})\rangle$ is simply the \textit{local} expectation value of the energy around vertex $v$. In this sense it differs from the free energy density, which cannot be expressed as a local expectation value. Again, the attuned reader may recognize \autoref{eq:Ev-ga} as the expression for the energy density within the BP approximation, which can be written in terms of the messages received by a ``factor'' (interaction node)~\cite{mezard2009information}.

We have claimed that the average energy density per vertex for $\gamma=\alpha=1$  matches that of the paramagnet (\autoref{eq:efs-para}), whereas the free energy densities differ.\footnote{Here, as throughout this work, the free energy for a given $(\gamma,\alpha)$ below its transition temperature is the intravalley free energy, or the ``internal'' free energy in the parlance of Ref.~\cite{Zdeborova2010}.} This is because the bulk energy is the average of a linear function of local marginals, which for $\gamma=\alpha=1$ exactly match the paramagnetic (replica-symmetric) solution~\cite{krzakala2007gibbs,Zdeborova2010}.  

Explicitly, for $\alpha=1$, if the distribution $\Q{}$ is symmetric in $m$, then the normalization factors $\mathcal{Z}_{\rm int}$ and $\mathcal{Z}_{\rm delta}$ are easily evaluated:
\begin{equation}\label{eq:Z-alpha1}
\mathcal{Z}_{\rm delta}(\beta, 1,\gamma) = 1/2, \quad \mathcal{Z}_{\rm int}(\beta, 1, \gamma) = Z(\gamma \beta)/2.
\end{equation} 
Plugging this into~\autoref{eq:Ev-ga} and writing out the denominator explicitly, with $\gamma=1$, 
\begin{align}\label{eq:e-1}
\varepsilon(\beta,1,1) &= \frac{\sum_{\spin_0,\vec{\spin}} E_v(\spin_0, \vec{\spin}) e^{-\beta E_v(\spin_0,\vec{\spin})}}{Z(\beta)/2} \int \prod_{i=0}^b \left(\frac{1+m_i \spin_i}{2}\right) \prod_{i=0}^b dQ_{\beta,1,1}(m_i)  \notag \\
&= -\left(\frac{1}{2}\right)^b\frac{\sum_{\sigma_0,\vec{\spin}} \partial(e^{-\beta E_v(\spin_0,\vec{\spin})})/\partial\beta}{Z(\beta)} = -\frac{\partial}{\partial \beta} [\log Z(\beta)] = \ec{para}(\beta).
\end{align}

For our analysis of high-energy valleys, we compute the energy density in a different way, described in \autoref{app:e-v2} of this Supplement.

\subsection{Moments of the spin}
To evaluate moments of the spin, we consider a final recursion step, with
\begin{equation}\label{eq:mbulk}
    m_{\rm bulk}(m_0,m_1) = \frac{\sum_{\spin=\pm 1} s (1+sm_0)(1+sm_1)}{\sum_{\spin=\pm 1} (1+m_0)(1+m_1)} = \frac{m_0 + m_1}{1 + m_0 m_1} = \left(\frac{\partial}{\partial m_0} + \frac{\partial}{\partial m_1}\right) \log(z_{\rm delta}(m_0,m_1)).
\end{equation}
This, again, can be expressed in the language of BP: $m_0, m_1$ parameterize the messages received by a given spin (a ``variable''), and $m_{\rm bulk}$ parameterizes the resulting marginal distribution (``belief'') on that spin~\cite{mezard2009information}.\footnote{A more familiar form of~\autoref{eq:mbulk} is $p_i(\spin_i) \propto \prod_{a\sim i} \nu_{a\rightarrow i}(\spin_i)$, where $p_i(\spin_i)$ is the marginal probability on variable $\spin_i$, and $\nu_{a\rightarrow i}$ is a message passed from factor $a$ to variable $i$. For the Tanner codes considered here, each variable has exactly two neighbors, so writing $\nu_{a\rightarrow i}(\spin_i) = (1 + m_{a\rightarrow i} \spin_i)/2$ recovers~\autoref{eq:mbulk}.}
The distribution of this bulk magnetization is then
\begin{equation}\label{eq:Q-bulk}
    \Q{\rm bulk}(m) = \frac{\int z_{\rm delta}(\mu_0,\mu_1)^\alpha \delta(m - m_{\rm bulk}(m_0,m_1)) d\Q{}(m_0)d\Q{}(m_1)}{\mathcal{Z}_{\rm delta}(\beta,\alpha,\gamma)}.
\end{equation}
A conventional probe of spin glass order is the second moment of the distribution, known as the Edwards-Anderson order parameter~\cite{Edwards1975}, which in our setting evaluates to:
\begin{equation}\label{eq:q-EA}
    q_{\rm EA}(\beta,\alpha,\gamma) \equiv \mathbbm{E}_{\vec{\spin}_\partial}\left[\langle \spin \rangle_{\partial}^2\right](\beta,\alpha,\gamma) = \mathbbm{E}_{\Q{\rm bulk}}\left[m^2\right].
\end{equation}
In the first equality, the expectation value is taken over different boundary configurations $\vec{\spin}_\partial$, and $\expval{\spin}_\partial$ is the expectation value of a bulk spin conditioned on that boundary configuration. Conditioning on the boundary is akin to evaluating expectation values within a single pure state (a single valley), as per the standard definition of the Edwards-Anderson order parameter ~\cite{parisi1983order,binder1986review}.\footnote{See also Ref.~\cite{chayes1986mean}, which studies~\autoref{eq:q-EA} in the Bethe lattice Ising spin glass for the particular case $\alpha=\gamma=0$ (quenched BCs). As emphasized in Refs.~\cite{mezard2001bethe,mezard2006reconstruction}, however, these boundary conditions are inappropriate for probing replica symmetry breaking on the corresponding random regular graph.} 

\section{Review of prior work}\label{app:prior}

\subsection{Tanner-Hamming model}
First we briefly review results from Ref.~\cite{Placke_glass}, in which we focused primarily on $\alpha=\gamma=1$ (``typical'' BCs) and $\alpha=\gamma=\infty$ (``codeword-polarized'' BCs). 

Under codeword-polarized boundary conditions, we only join together compatible branches, i.e., only the branches with the largest $z(\vec{m},\beta)$ contribute. Thus the distribution $Q_{\beta,\infty,\infty}$ (or more generally, $Q_{\beta,\alpha,\infty}$ for any $\alpha > 0$) is described at all depths by a single scalar $m$, converging to distribution
\begin{equation}\label{eq:ferro-fp}
Q_{\beta,\infty,\infty}(m) = \frac{1}{2} \left[\delta(m-m_*(x)) + \delta(m+m_*(x))\right]. 
\end{equation}
where $m_*$ is a fixed point of the recursion relation~\autoref{eq:fm-ferro}. The paramagnet ($m_*=0$) is always a stable fixed point, but for $x < x_{\rm mem} \equiv 1/25$, any initial condition with large enough $|m_0|$ will flow towards a nontrivial fixed point:
\begin{equation}\label{eq:ferro-m}
m_{\rm ferro}(x) = \sqrt{\frac{1}{3} \left(\sqrt{\frac{1-25 x}{1-x}}+2\right)}.
\end{equation}
This ``ferromagnetic'' fixed point is stable for $x>x_{\rm mem}$ and separated from the paramagnet by an unstable fixed point $0 < m_{\rm unstable}(x) < m_{\rm ferro}(x)$. The existence of a nontrivial fixed point indicates \textit{weak} clustering: the open-boundary (paramagnetic) Gibbs state is not unique, but it still dominates the landscape, and only fine-tuned boundary conditions can access stable valleys. At $x=x_{\rm mem}$, $m_{\rm unstable}$ and $m_{\rm ferro}$ collide and become marginal, signaling a canonical instability within the valleys around codewords. The cusp in $m_{\rm ferro}(x)$ is accompanied by a cusp in the energy, as we discuss below.

Under typical BCs, we use population dynamics together with a low-temperature expansion to verify the existence of a spin glass phase. This phase has \textit{strong} clustering: no pure state dominates the weight, and indeed, the configurational entropy density \textit{increases} with $T$ at low temperatures. The mutual information between the root and leaves, as well as even moments of the bulk magnetization distribution, are nonzero within the spin glass phase, then jump discontinuously to zero at $T=\Tglass$.

In Ref.~\cite{Placke_glass}, we also briefly touched on two other BC ensembles: $\gamma=\alpha=0$ and $\gamma=\alpha=2$. Using population dynamics, we showed that the former ensemble is paramagnetic at all finite temperatures: $Q_{\beta, 0, 0}^*(m) = \delta(m)$. In fact, the zero-temperature state is also a paramagnet, due to entropic fluctuations. In contrast, for any fixed $\gamma, \alpha > 0$, the $T=0$ state is ordered and ferromagnetic because the factor of $z(\vec{\mu},\gamma\beta/\alpha\rightarrow\infty)^\alpha$ in~\autoref{eq:family} selects only the lowest energy branches; we will explore the critical temperature at small $\gamma=\alpha$ in~\autoref{app:ga-fam} below.

The choice $\gamma=\alpha=2$, yields an ``annealed average'' of the Edwards-Anderson order parameter, and its transition temperature $T_c(2,2)=0.2522...$ is exactly solvable via recursion on a two-copy tree. This will serve as a useful cross-check on our numerics and on the low-temperature expansion for the bulk magnetization.

The discerning reader will note that the energy scale in the present work differs from Ref.~\cite{Placke_glass}. This comes about for the reason already noted in footnote~\onlinecite{tanner}. Typically, one maps a parity check matrix $H$ onto a Hamiltonian $\mathcal{H}$ as $\mathcal{H}(\vec{\sigma}) = |H \mathbf{b}|$, where $\spin_i = 1-2 b_i$. However, if we took this convention for the local parity check matrix $H_{\rm L}$, we would be distinguishing between local configurations that violate different numbers of local checks. Instead, we only want to distinguish between codewords and non-codewords. In Ref.~\cite{Placke_glass}, we achieved this by replacing $H_{\rm L}$ with its symmetrized (and locally redundant) version of $H^{(\rm symm)}_{\rm L}$, whose rows are all $2^{\nL-\kL}-1$ nontrivial checks of the Hamming code. Then, violating any number of local checks incurs an energy penalty of $2^{\nL-\kL-1}$.  In this work, we instead assign an energy penalty of 1 to each violated node, so all units of energy are modified by a factor of $2^{\nL-\kL-1}=4$ relative to Ref.~\cite{Placke_glass}.

\subsection{Connection to the broader literature}
Though presented in a different style, the methods laid out in the previous two sections of this Supplement share many ingredients with decades of work on the dynamical glass transition in the Ising model~\cite{mezard2001bethe,Mezard2003,mezard2006reconstruction} and its $p$-spin variant on random regular graphs~\cite{Barrat1999,Ricci2001,mezard2003two_solutions,montanari2003nature,montanari2004cooling_schedule,Bernaschi2021,franz2001ferromagnet,Franz2001exact,Bellitti2021,Krzakala2010,Zdeborova2010,franz2002dynamic,montanari2001glassy}. Before proceeding, let us comment now on the similarities and differences between the present work and the literature, and the extent to which our model can be considered an unfrustrated, non-random spin class.

A well-studied model in the community is the \textit{diluted p-spin model}, with the Hamiltonian~\cite{gallager1960thesis,gallager1962low,Barrat1999}
\begin{equation}\label{eq:p-spin}
    \mathcal{H}(\vec{\sigma}) = -\sum_{a} J_a \prod_{i \in \partial a} \spin_i,
\end{equation}
Here $a$ labels the factors (interactions), $J_a$ the coupling strength, and $\partial a$ is the set of $p$ variables (spins) involved in that interaction. This can be represented as a bipartite factor graph $G$, or Tanner graph, consisting of interaction nodes $\{a\}$ and spin nodes $\{i\}$~\cite{montanari2001glassy}. If the model is LDPC ($\deg(a) = p=O(1) \forall a$, and $\deg(i) = O(1) \forall i$), then the factor graph is sparse. Often, one chooses $G$ to be a random biregular graph, i.e. each spin is involved in exactly $c$ checks~\cite{montanari2001glassy}; alternatively, one can allow the spin degree to fluctuate while fixing the ratio $\gamma = N_a / N_i$~\cite{Barrat1999,Ricci2001,franz2001ferromagnet,Franz2001exact,mezard2003two_solutions}. 

In the theoretical computer science literature, \autoref{eq:p-spin} defines the constraint satisfaction problem XOR-SAT~\cite{ricci_tersenghi2010xorsat}.  While early studies considered random couplings $J_a = \pm 1$, much of the glassy phenomenology survives to the uniform case $J_a=1$~\cite{Barrat1999,Ricci2001,franz2001ferromagnet,mezard2003two_solutions}. \autoref{eq:p-spin} then has at least one ground state, the all up state, called a ``planted'' solution~\cite{Ricci2001}. In the coding theory literature, \autoref{eq:p-spin} with $J_a=1$ and $c < p$ defines a \textit{Gallager code}~\cite{gallager1960thesis,gallager1962low}, with code rate $1 - c/p$.\footnote{Like our Hamming-Tanner code, Gallager codes are redundancy-free with high probability; unlike the Hamming-Tanner code, this property has been rigorously proven~\cite{richardson2008modern}.}

While the Tanner-Hamming code (and Tanner-Ising models more broadly~\cite{Placke_glass}) and diluted $p$-spin models both present examples of sparse, unfrustrated spin glasses, our work differs from the literature in a few notable ways. First, analytical and numerical studies of XOR-SAT have largely focused on the \textit{overconstrained} case $c \geq p$~\cite{franz2001ferromagnet,Bernaschi2021,Bellitti2021}, which therefore does not have a finite code rate. In particular, the extensive redundancies in models with $c>p$ mean that the partition function does not factorize, and the paramagnetic fixed point is thermodynamically stable only at high temperatures. Below the dynamical glass transition is a \textit{static} 1RSB transition, to a phase in which a few pure states dominate the Gibbs measure and the configurational entropy density vanishes~\cite{franz2001ferromagnet}. In contrast, the extensive code rate and lack of redundancies in our models rule out static 1RSB.

symmetry breaking.

The $c<p$ parameter regime relevant to error correction is more closely related to our model. Indeed, for sufficiently large $p$, Gallager codes meet the criteria of the proof in Ref.~\cite{Placke_glass} for an incongruent shattering phase at low temperatures. Earlier studies have applied the cavity method and 1RSB formalism to these models~\cite{montanari2001glassy,franz2002dynamic}\footnote{It should be noted, though, that these works are more concerned with the statistical physics of active error correction, not finite-temperature boundary reconstruction.}, and we consider it likely that the additional phenomena discussed in this work, such as the extended phase diagram of high-energy valleys and the separation of microcanonical and canonical instabilities, also appear in these models.

Both Gallager codes and the Tanner-Hamming model on locally tree-like graphs are manifestly unfrustrated in the sense that every interaction can be independently satisfied. While one can view the boundary conditions on the tree as introducing frustration~\cite{mezard2006reconstruction}, the glassy phase is not merely an artifact of boundary conditions: it is evident from simulations on closed graphs, without requiring sign disorder on the couplings. In this sense, diluted $(p>2)$-spin glasses, Tanner-Hamming codes, and LDPC expander codes more broadly differ from the ferromagnetic Ising model ($p=2$), where the spin glass phase on the Bethe lattice is unstable towards the ferromagnet when one closes the boundaries~\cite{franz2001ferromagnet,mezard2006reconstruction}.

What about disorder? We have called our model a ``non-random spin glass'', and while there is indeed no quenched disorder in the couplings, one might rightfully object that the model is defined on a \textit{random} regular graph (RRG). 

We argue that this randomness is mild, in the sense discussed in Ref.~\cite{franz2001ferromagnet}: the graph is deterministic out to the size of the smallest loop, which diverges in the thermodynamic limit. Moreover, while the locally tree-like structure of the RRG is essential to the recursive techniques we employ, spin glass order in LDPC expander codes is a far more general phenomenon~\cite{Placke_glass}. The essential ingredient is not the absence of loops, but a subextensive number of redundancies together with sufficiently strong code expansion. A numerical study of Tanner-Hamming codes on the hyperbolic lattice provides a concrete realization of a model without any frustration or randomness, with much the same phenomenology as the model on a random regular graph~\cite{Placke_glass}. While Ref.~\cite{Placke_glass} demonstrated weak and strong clustering via heating and cooling simulations, it  will be interesting to study the landscape of valleys beyond these two transitions, analogously to the case study undertaken here for the locally tree-like model. We leave this avenue to future work, proceeding now instead to the derivation of this Letter's main results.

\section{Low-temperature expansion for $(\alpha,\gamma)$ families}\label{app:low-T}
In this section we derive the low-temperature expansion for the fixed-point distribution in the $(\beta,\alpha,\gamma)$ families. To distinguish between ``thermal'' contributions to the free energy (which appear at $O(x^2)$ and above) and ``defect'' contributions (appearing at $O(x^{\gamma})$ and above), we keep terms up to and including $O(x^{1+\gamma}), O(x^{\gamma(1+1/\alpha)}, O(x^3)$.

\subsection{Fixed point distribution at root}
To find $\Q{*}$, we initialize the recursion relation from the ferromagnetic fixed point (\autoref{eq:ferro-m}), expanding $m_* =m_{\rm ferro}(x)$ up to $O(x^3)$:
\begin{equation}
m_*(x) = 1 - 2x - 16x^2 - 202x^3 - O(x^4).
\end{equation}

The distribution converges after a finite number of iterations:
\begin{description}
\item[Step 1]A defect in the node directly above the root (e.g. one spin $m=-m_*$, the rest $m=+m_*$) produces 
\begin{equation}\label{eq:m1}
    m=m_1=0
\end{equation}
on the root.
\item[Step 2] Setting all inputs $m=m_*$, except one $m=m_1$, creates 
\begin{equation}\label{eq:m2}
m=m_2=1-6x-20x^2-294x^3
\end{equation}
at the root. This term comes from a defect spin at distance 2 from the root. Since the distance is 2, it differs from $m_*$ only at order $O(x)$ and higher. 
\item[Step 3] Setting all inputs $m=m_*$ except one $m=m_2$ creates
\begin{equation}\label{eq:m3}
m_3=1-2x-24x^2-282x^3.
\end{equation} Now the distance from a defect is 3, and the new term differs from $m_*$ only at order $O(x^2)$ and higher. 
\item[Step 4] Setting all inputs $m=m_*$ except one $m=m_3$ creates 
\begin{equation}\label{eq:m4}m_4=1-2x-16x^2-202x^3.
\end{equation} This differs from $m_*$ only at order $O(x^3)$ and higher.
\item[Step 5] No new terms appear here to $O(x^3)$, because a defect more than 4 steps away will only change the root spin by $O(x^4)$ or higher. To wit, plugging in $m=m_*$ on 5 inputs and $m=m_4$ on the sixth input yields $m=m_* + O(x^4)$ on the root.
\end{description}

Altogether, the distribution converges to
\begin{align}\label{eq:series-gamma-alpha2}
\Q{*}(m) = \sum_{\spin=\pm 1} \Bigg[&\frac{1}{2}\left(1-3 \cdot 2^\alpha x^\gamma(2^\alpha + 516 + \alpha x^{\gamma/\alpha} (7 \cdot 2^\alpha + 4668))\right) \delta\left(m - \spin m_*\right) \notag \\
+ &648 \cdot 2^\alpha x^\gamma (1 + 9\alpha x^{\gamma/\alpha})\delta\left(m - \spin m_4\right) +108 \cdot 2^\alpha x^\gamma (1 + 9\alpha x^{\gamma/\alpha}) \delta\left(m-\spin m_3\right) \notag \\
+& 18 \cdot 2^\alpha x^\gamma (1 + 11\alpha x^{\gamma/\alpha}) \delta\left(m - \spin m_2\right)\Bigg] + 3 \cdot 4^\alpha x^\gamma (1 + 7 \alpha x^{\gamma/\alpha}) \delta(m) + O\left(x^4,x^{2\gamma},x^{\gamma(1+2/\alpha)}\right).
\end{align}
The weights $Q_{\pm m_i}$ on different delta-function peaks arise by noting that the ``defect-free'' term (all inputs $m=m_*$) has
\begin{equation}\label{eq:no-defect}
z(\mu_*, \gamma\beta/\alpha)^\alpha = 1 - 5\alpha x^{\gamma/\alpha} - O(x^{2\gamma/\alpha})
\end{equation}
The product $z(\mu_*,\gamma\beta/\alpha)^\alpha (Q_{m_*})^b$ is the leading contribution to $\mathcal{Z}_{\rm iter}(\beta,\alpha,\gamma)$ in the denominator of the recursion relation~\autoref{eq:family-normalized}. The defect terms will contribute at $O(x^\gamma)$ and above to the denominator, so can be neglected in the denominator when computing $Q_{\pm m_i}$. We then obtain the weights recursively through
\begin{subequations}
\begin{align}
    Q_0 &= \frac{3 z((\mu_*,\mu_*,\mu_*,\mu_*,\mu_*,-\mu_*),\gamma\beta/\alpha)^\alpha}{z(\mu_*, \gamma\beta/\alpha)^\alpha} = 3 \cdot 4^\alpha x^\gamma (1 + 7\alpha x^{\gamma/\alpha}) + O\left(x^{2\gamma},x^{\gamma(1+2/\alpha)}\right).\\
    Q_{\pm m_i} &= \frac{6 z((\mu_*,\mu_*,\mu_*,\mu_*,\mu_*,\mu_{i-1}),\gamma\beta/\alpha)^\alpha}{z(\mu_*, \gamma\beta/\alpha)^\alpha} Q_{\pm m_{i-1}}.
\end{align}
\end{subequations}
Finally, the coefficient $Q_{\pm m_*}$ is determined by requiring that all weights sum to 1.

\subsection{Free energy, energy, defect density}
Substituting~\autoref{eq:series-gamma-alpha2} into ~\autoref{eq:Y-bulk},~\autoref{eq:Y-delta},~\autoref{eq:Y-int}, and~\autoref{eq:beta-f} yields

\begin{align}\label{eq:f-gamma-alpha-series}
\beta f(x,z,\alpha) &=  -\frac{7}{2} x^2 -28 x^3 - \frac{7z}{2}  \left[\log(2x) + 9 (x + \frac{\alpha}{2} z^{1/\alpha} \log(2x))\right],
\end{align}
where we have defined $z=2^\alpha x^\gamma$ to remove the explicit dependence on $\gamma$.\footnote{Not to be confused with the local partition function $z(\vec{m})$.} Similarly, from~\autoref{eq:e-gamma-alpha} we obtain
\begin{align}\label{eq:e-gamma-alpha-series}
\varepsilon(x, z,\alpha) &= 7x^2 + 84x^3 + \frac{7z}{2} \left[1 + 9 (x + \frac{\alpha}{2} z^{1/\alpha})\right]
\end{align}

The similarity of~\autoref{eq:f-gamma-alpha-series} and~\autoref{eq:e-gamma-alpha-series} leads us to define a ``defect density''
\begin{equation}
\tilde{z} =  z \left(1 + \frac{9}{2} \alpha z^{1/\alpha}\right),
\end{equation}
reducing $\beta f, \varepsilon$ to functions of only two variables $x, \tilde{z}$. For easy reference, we recall \autoref{eq:f-defect} and~\autoref{eq:e-defect} from the main text:
\begin{equation}\label{eq:defect-both}
    \beta f(x,\tilde{z}) = -\frac{7}{2} x^2 -28 x^3 - \frac{7}{2} \tilde{z} \left[\log(2x) + 9x + ...\right], \qquad
\varepsilon(x, \tilde{z}) = 7x^2 + 84x^3 + \frac{7}{2} \tilde{z}(1+9x+...).  
\end{equation}
Thus, at this order, the parameters $\gamma,\alpha$ do not provide independent tuning knobs but together control the defect density at fixed $x$.

Probing the defect correlations would require expanding to order $x^{2\gamma}$. However, the series expansion does not appear to converge with a finite number of terms at this order, so we have chosen not to pursue this avenue further.

To gain some physical intuition for~\autoref{eq:defect-both}, suppose $\gamma < 1$ so that the leading terms are the defect terms $\sim\tilde{z}$ and $\sim\tilde{z}x$.  Each defect is a flippable spin, which has adjacent to it one violated node in the ground states of the defect. The defect has two ground states: flipping the spin moves the violated node, keeping the energy at 1. There are additionally 18 excited states at energy 2, obtained by either flipping one of the other spins adjacent to the violated node (12 possibilities) or flipping two neighboring spins in such a way that the violated node is now satisfied, but making two new violations in the process (6 possibilities). Thus, the partition function per node in the bulk is $(2x+18x^2)^{\tilde{z}}$ at this order, in agreement with $\beta f(x,\tilde{z})$ given above. At this order, the defect has probability $(1-9x)$ of having energy 1 and probability $9x$ of having energy 2, which agrees with the $\varepsilon(x,\tilde{z})$ given above.

\subsection{Configurational entropy density}

The configurational entropy density along the line $\alpha=\gamma$ can likewise be expressed in terms of the defect density:
\begin{align}\label{eq:s-conf-gamma}
    s_{\rm conf}(\beta, \alpha=\gamma) = \log(2)/2 + \frac{7}{2} (2x)^\gamma[1-\gamma(1+9 \gamma x)\log(2x)] = \log(2)/2 + \frac{7}{2} \tilde{z}(1-\log\tilde{z}) + O(x^4,x^{2\gamma}, x^{\gamma+1}).
\end{align}
This reduces to~\autoref{eq:s-conf-x} of the main text upon substituting $\gamma=1$.
\subsection{Bulk magnetization}
Plugging the low-temperature expansion for $\Q{}$ into~\autoref{eq:Q-bulk} yields
\begin{align}\label{eq:bulk-Q-low}
    \Q{\rm bulk}(m)& = \sum_{\spin=\pm 1}\frac{1}{2}(1-3109 \tilde{z}) \delta(m-\spin m_{\rm bulk}(m_*,m_*)) + \sum_{j=1}^4 \sum_{\spin=\pm 1} 6^j \tilde{z} \delta(m - \spin m_{\rm bulk}(m_*,m_i)) \notag \\
    &+ \tilde{z} \delta(m) + O\left(x^4,x^{2\gamma}, x^{\gamma+2},x^{\gamma(1+2/\alpha)}\right),
\end{align}
where $m_j$ is the root magnetization associated with a distance $j$ defect, as defined in~\autoref{eq:m1}-\autoref{eq:m4}.\footnote{To the order we have carried the calculation, $m_{\rm bulk}(m_*,m_*)=m_{\rm bulk}(m_*,m_4)$, but we choose to keep the terms separate to illuminate the structure in the delta function weights.}

\autoref{eq:bulk-Q-low} further justifies the designation of $\tilde{z}$ as the defect density. When two rooted trees with root magnetizations of opposite sign are joined at the root, the merged root spin hosts a defect. Accordingly, the weight of the delta function $\delta(m)$ is $\tilde{z}$. Similarly, merging one defect-free rooted tree ($m_{root} = m_*$) with a rooted tree hosting a defect at distance $j$ from the root ($m_{root} = m_j$) produces a weight proportional to $\tilde{z}$. A single defect at distance $j$ is diluted by a factor of $2 \times b^j = 2\times 6^j$, which is precisely offset by the weight on the delta function.

To the same order, the Edwards-Anderson order parameter is
\begin{equation}\label{eq:qEA}
q_{\rm EA}(x,\tilde{z}) = 1-4x^2-72x^3-\tilde{z}-48x\tilde{z}.
\end{equation}

Again, we can gain physical intuition for ~\autoref{eq:qEA} by considering the case $\gamma < 1$, where the leading terms are the defect terms $\sim \tilde{z}, \sim \tilde{z} x$. Consider an arbitrary bulk spin $e$. With probability $\tilde{z}$, $e$ hosts a defect and is flippable, so $m_{\rm bulk}^2 = 0$. With probability $12\tilde{z} - O(\tilde{z}^2)$, the spin is not flippable in the ground state, but one of its neighbors is, and there are two excited states in which $e$ is flipped, at the cost of raising the energy by 1, so that $m_{\rm bulk}^2 = 1 -4x + O(x^2)$. With the remaining probability (defects at distance $>1$), any configuration in which spin $e$ is flipped raises the energy by 2. Therefore, at this order, $q_{EA} = (1-13\tilde{z}) + 12\tilde{z}(1-4x)$, in agreement with~\autoref{eq:qEA}. 

A further sanity check comes from considering $\alpha=\gamma=2$, where the second moment (but not the full distribution) can be computed exactly~\cite{Placke_glass}:
\begin{equation}
    q_{\rm EA}(x,2,2) = \frac{2 (1-x) \sqrt{6-\frac{3 \sqrt{1-x (143 x+50)}}{x-1}}}{-5 x+\sqrt{1-x (143 x+50)}+5} = 1 - 8x^2 - 336x^3 - O(x^4),
\end{equation}
in agreement with~\autoref{eq:qEA}.

\subsection{Supporting numerics}\label{app:ga-fam}

\begin{figure}[t]
\centering
\includegraphics[width=\linewidth]{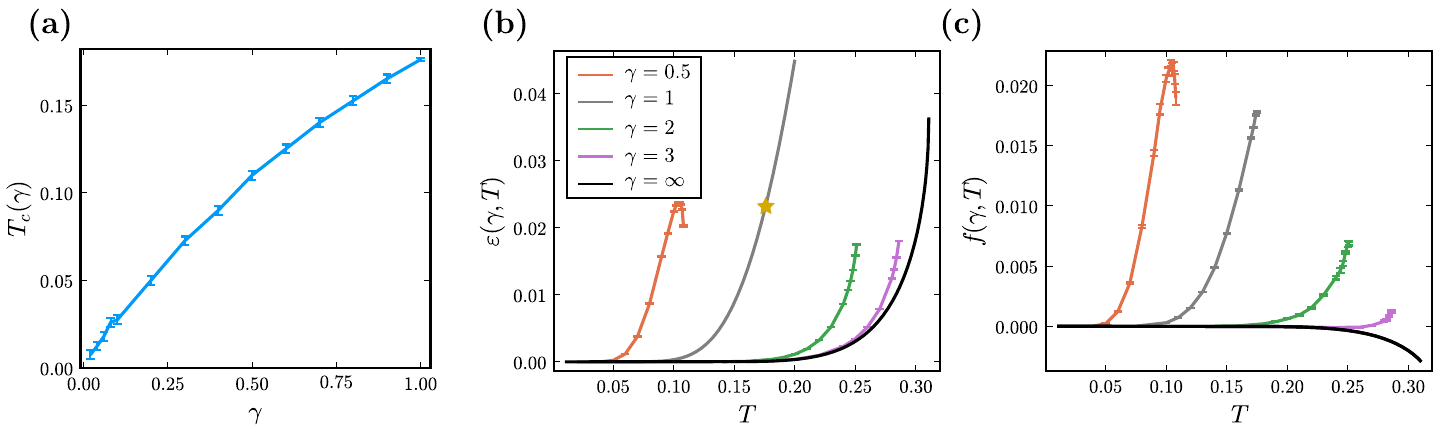}
\caption{Phase transitions in $\gamma=\alpha$ families. (a) $T_c(\gamma)$ for small $\gamma$. (b) Energy density and (c) free energy density within the ordered phase, for $\gamma=0.5, 2, 3$ (using population dynamics), and $\gamma=\infty$ (analytic calculation). In (b), paramagnetic energy density is indicated by the gray curve; this is also the energy density for $\gamma=1$, whose critical temperature is marked by a yellow star. In (c), the free energy density for $\gamma=1$ was evaluated using population dynamics.\label{fig:e-f-sg}}
\end{figure}

Going beyond the low-temperature expansion, we first explore the trend in $T_c(\gamma)$ along the line $\gamma=\alpha$. \autoref{fig:e-f-sg}a shows the critical temperature $T_c(\gamma)$ down to $\gamma = 0.02$, where $T_c(0.02) = 0.0075 \pm 0.0025$. The data are consistent with $T_c(\gamma)$ going continuously to zero as $\gamma \rightarrow 0^+$. Two other possibilities remain. (1) There may exist $0 < \gamma_c < 0.02$ such that for $\gamma < \gamma_c$, there is no ordering at any finite temperature. (2) It may be that $\lim_{\gamma\rightarrow 0^+} T_c(\gamma) > 0$, but even if so, the monotonicity of the critical temperature with $\gamma$ bounds the magnitude of the discontinuity: $T_c(\gamma \rightarrow 0^+) \leq T_c(0.02)$.

We noted in the main text that the direction and existence of a cusp at $T_c$ depends on $(\gamma,\alpha)$. A few representative examples appear in \autoref{fig:e-f-sg}b, all with $\gamma=\alpha$: $\gamma=0.5,1,2,3,\infty$. All curves are shown only in the ordered phase, with the exception of $\varepsilon(1,T)$, which coincides with the paramagnetic energy density at all temperatures.
At $\gamma=0.5$, the cusps in both $f$ and $\varepsilon$ are downward, indicating that the instability is dominated by the entropic dilution of defects. At $\gamma=2$ and $\gamma=3$, both quantities exhibit an upwards cusp, indicating that the instability is dominated by thermal excitations.\footnote{Though the cusps in $f(2,T)$ and $f(3,T)$ are less evident, this is due in part to the fact that we do not gather data up to precisely $T_c(\gamma)$, as the population dynamics method becomes less reliable as the transition is approached. For example, for $\gamma=2$ we only show the (free) energy density up to $T=0.251$, vs. the analytically computed critical temperature $T_c(2) = 0.2522...$.} $\gamma=\infty$ is a special case: the bulk energy density has an upwards cusp (defects are forbidden, so the instability is solely due to thermal excitations within a valley), but the free energy can only have a jump discontinuity, since $s = -\partial f/\partial T$ is finite.

\begin{figure}[t]
\includegraphics[width=0.8\linewidth]{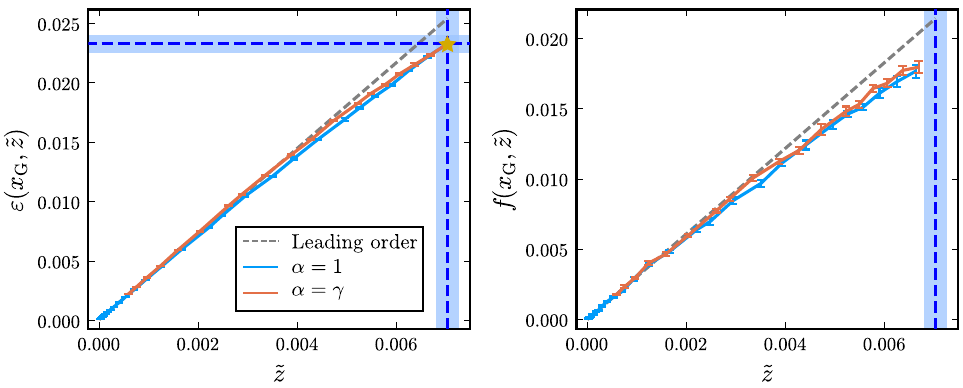}
\caption{Energy density (left) and free energy density (right) at $T=\Tglass = 0.176$, plotted vs. the effective defect density $\tilde{z}$ (\autoref{eq:defect-density}) for two families: $(\alpha,\gamma)=(1,\gamma)$ (blue) and $(\alpha,\gamma)=(\gamma, \gamma)$ (orange), with $\gamma$ approaching 1 from above for both families. The gray dashed line is the series expansion at leading order in $\tilde{z}$ (\autoref{eq:e-defect} and~\autoref{eq:f-defect}). Vertical blue dashed lines indicate the defect density for $\alpha=\gamma=1, T=\Tglass$. In the left panel, the horizontal blue dashed line indicates the paramagnetic energy density at $T=\Tglass$.
\label{fig:e-f-sg1}}
\end{figure}

The most subtle case is $\gamma=\alpha=1$, for which neither $f$ nor $\varepsilon$ have cusps.\footnote{As noted in the main text, the absence of a cusp in $f$ follows from the assumption that $s(\varepsilon)$ is a smooth function of $\varepsilon$ within a valley. Numerical data support this for a range of $(\alpha,\gamma)$, and indeed $f(1,T)$ plotted up to $T=0.175$ in~\autoref{fig:e-f-sg}c does not exhibit a cusp, although the curve could technically sharpen in the narrow interval $[0.175,\Tglass]$.} A natural question is whether there is a cusp as the spin glass transition is approached from a different direction. We explore this by fixing $x=x_{\rm G}$ and approaching $\gamma=\alpha=1$ along two curves: $\gamma=\alpha \rightarrow 1^+$ and $\alpha=1,\gamma\rightarrow 1^+$. The resulting energy density and free energy density are plotted in~\autoref{fig:e-f-sg1}. We do not see any indication of a singularity. The gray dashed lines show the leading order series expansion for $\varepsilon(x,\tilde{z})$ and $f(x,\tilde{z})$; as expected, the numerical curves deviate from the leading order calculation as $\tilde{z}$ increases. There is also a slight discrepancy between the two numerical curves for the energy, indicating additional dependence on $(\alpha,\gamma)$ beyond that captured by the defect density up to order $O(x^{\gamma(1+1/\alpha)})$.  

\section{Stabilizing higher energies}\label{app:fine-tuned}
As in the main text, we consider 3 types of nodes: type 0, 1, and 2. We continue to focus on the Tanner-Hamming code, but where possible comment on how the method would generalize to other local codes.

\subsection{General framework}
First, let's fix notation. Nodes of type 1 and 2 come in two flavors, with positive and negative magnetization, while type 0 nodes have a root magnetization of 0. We will use Greek letters to denote the signed node types ($\alpha=0,\pm 1, \pm 2$) and Latin letters to denote the unsigned types ($j=0,1,2$). The conditional magnetization at the root of a type $j$ node is $\pm m_j$.

We denote the input branches to a type $\alpha$ node by a length-$b$ vector $\vec{c}_\alpha$, where $c_{\alpha,i} \in \{0,\pm 1,\pm 2\}$ and $\vec{c}_{-\alpha} = -\vec{c}^{\alpha}$.
We then define the $3\times 3$ matrix $\mathbf{N}$, where $N_{ij}$ denotes the number of type $i$ branches (inputs) used to make a type $j$ root (output). That is, 
\begin{equation}
    N_{ij} = \sum_{k=1}^b \delta_{|c_{j,k}|,i}.
\end{equation}
The matrix $\mathbf{N}$ uniquely determines the family of valleys once we impose that type 0 nodes host a defect, and type 1 and type 2 nodes are frozen at $T=0$.\footnote{That is, all choices of $\vec{c}_{\alpha}$ consistent with $\mathbf{N}_{\cdot,|\alpha|}$ give rise to the same recursion relations for the energy and magnetization, once we exclude the ordering and signs on branches that would force a defect onto a type 1 or type 2 node, or that would not force a defect onto a type 0 node.} Each column sums to $b=6$, and to prevent the entropy from diverging at low temperatures (thereby destabilizing the valley), we impose additional constraints on the entries of the matrix. 

Given $\mathbf{N}$, we can solve for the density of $d_0, d_1, d_2$ branches using
\begin{equation}
\mathbf{N} \vec{d} = b \vec{d}.
\end{equation}

Let $E_\alpha(\spin_0) \in \{0, 1\}$ denote the minimum energy of a type $\alpha$ node with the output (root) spin fixed to $\spin_0$. At $T=0$, any inputs of type $\pm 1$ or $\pm 2$ are frozen to $\sigma_0 = \pm 1$, while inputs of type 0 are free. Let $\Omega_\alpha(\spin_0)$ denote the degeneracy of minimum-energy states with the root fixed. 

The three node types can be distinguished along two axes: energy and entropy. 

Types 1 and 2 are frozen at zero temperature. A positive node has $E_{\alpha}(+1)=0, E_{\alpha}(-1)=1$, and vice versa for a negative node. The ground state of the root is unique: $\Omega_{\pm j}(\pm 1) = 1$.

Type 0 nodes contain a defect: the frozen spin states on the input branches are chosen with signs that are incompatible with a codeword regardless of the state on the output, i.e. $E_{0}(1)=E_{0}(-1)=1$.

Along the entropy axis, type 0 and type 2 nodes both have no entropy difference at the root, $\Omega_{\alpha}(1)=\Omega_{\alpha}(-1)$. For type 0 nodes, this ensures that the root magnetization is zero at all temperatures, including $T=0$. Type 1 nodes, on the other hand, have an entropy difference at the root: $\Omega_{\pm 1}(\mp 1)=2^{N_{01}}$, since if the output spin is flipped (thereby exciting the root node), all of the type 0 inputs become flippable. Thus, paying the energy cost of an excited root node provides an entropy gain further up the tree.

At $T=0$, a naive prediction for the energy density is the total defect density $d_0$. In practice, however, the defects travel towards the leaves to maximize entropy, so the bulk energy density is reduced from this prediction by a multiplicative factor.

\subsection{The highest-energy example}
Let us begin with an example, choosing $\mathbf{N}$ to obtain the highest energy density at $T=0$ while ensuring stability to nonzero temperatures:
\begin{equation}\label{eq:N-max}
\mathbf{N}_{\rm max} = \begin{pmatrix}
2 & 3 & 0 \\
0 & 0 & 6 \\
4 & 3 & 0 \\
\end{pmatrix} \Rightarrow \vec{d}_{\rm max} = \begin{pmatrix}
    3/11 \\
    4/11 \\
    4/11
    \end{pmatrix}.
\end{equation}

Thus, the energy density at zero temperature is upper bounded by $d_0=3/11$. However, the true energy density at $T=0$ is only $\varepsilon(0) = 7/44$. The discrepancy is explained by the upward mobility of the defects: a defect at the root can move up along a branch if, by flipping that branch to an unfavorable spin, the root can be satisfied. When $N_{10}=0$, at $T=0$, a defect hosted on a type 0 node can move up to a type 2 node in the layer above with total probability $p=1/2$, but cannot move any higher. The bulk defect density is therefore reduced by a factor $(b+1) p / b = 7/12$. At $T>0$, there is a competition between the enhanced mobility of defects up the tree (which reduces the bulk defect density) and thermal excitations (which raise the bulk energy density above the defect density). The energy density as a function of temperature is determined using the method in~\autoref{app:e-v2} below. For $\mathbf{N} = \mathbf{N}_{\rm max}$, $\varepsilon(T)$ is a monotone decreasing function, with a downward cusp at $T=T_c$. It is drawn as the topmost orange curve in~\autoref{fig:e-together}.

To determine the critical temperature, we need to solve the coupled recursion relations for $m_1$ and $m_2$. A concrete choice of input vectors $\vec{c}$ consistent with~\autoref{eq:N-max} is
\begin{equation}
    \vec{c}_{0} = \begin{pmatrix} 2 & 2 & -2 & 0 & 2 & 0 \end{pmatrix}, \quad \vec{c}_1 = \begin{pmatrix} 2 & 0 & 2 & 0 & 0 & 2 \end{pmatrix}, \quad \vec{c}_2 = \begin{pmatrix} 1 & 1 & 1 & 1 & 1 & 1 \end{pmatrix}.
\end{equation}
Substituting into~\autoref{eq:f} yields
\begin{align}\label{eq:m-high}
m_1' = g((m_2,0,m_2,0,0,m_2),x) = \frac{m_2^3(1-x)}{1 + 7x}, \qquad
m_2' = g(m_1,x).
\end{align}
\autoref{eq:m-high} admits nontrivial fixed points up to $x_c=0.003289...$, corresponding to a critical temperature of $T_c\approx 0.175$. 

\subsection{Families of $\mathbf{N}$}
Generalizing now to other $\mathbf{N}$, the constraints on each column of the matrix are as follows:
\begin{description}
\item[First column ($m_0=0$)] To force a defect, $N_{00} \leq 2$, or more generally, $b-N_{00}\geq d_{\rm L}^\perp$, where $d_{\rm L}^\perp$ is the distance of the code dual to $\mathcal{C}$.\footnote{\label{footnote:defect}To force a defect, we must freeze a subset of spins to a configuration that is incompatible with all codewords, which means that an element of the row space of $H_{\rm L}$ must be supported entirely on this subset. Linear combinations of parity checks are precisely the codewords of the dual code~\cite{guruswami2019essential}.} Varying $(N_{10}, N_{20})$ does not change the recursion relations for $m_1, m_2$, as long as the composition of branches forcing the defect can be chosen so that the entropy difference at the root is zero. If $N_{00}=n$, any combination of $(N_{10}, N_{20})$ summing to $b-n$ is allowed. These different combinations can lead to different energy densities. We find empirically that choose all remaining branches to be type 2, i.e. $N_{20}=b-n$, leads to the highest energy density. 
\item[Second column ($m_1$)] To freeze the root at zero temperature, $N_{01} \leq 3$, or more generally, $N_{01} \leq b+1-d^\perp_{\rm L}$.\footnote{More precisely: A subset $S$ is said to be a \textit{recovery set} for the $i$th bit of a code if the restriction of a codeword to $S$ uniquely determines the state of $i$~\cite{Gopalan2012,guruswami2019essential}. In other words, given any codeword $\vec{\spin}$, there are no codewords compatible with freezing $S$ to match $\vec{\spin}$ and freezing $i$ to differ from $\vec{\spin}$. Per footnote~\ref{footnote:defect}, this means that there is a (linear combination of) parity checks that touches spin $i$ and is contained on $S  \cup \{i\}$~\cite{Guruswami2019}.} If $N_{01}=0$, then combined with the constraint $N_{02}=0$ discussed below, this would force $d_0 = 0$, so no defects are made and we recover codeword-polarized BCs. If $N_{01}>0$, inputs of type 2 must be used for any spins that would become flippable if the root is excited. For $N_{01} = 1$, this constrains one of the remaining five branches, so $N_{21} \geq 1, N_{11}\leq 4$ (five total choices). For $N_{01}=2$, three of the four remaining branches are constrained, so $N_{21} \geq 3, N_{11}\leq 1$ (two choices). For $N_{01}=3$, all three remaining branches are constrained, so the only choice is $N_{21}=3, N_{11} = 0$.
\item[Third column ($m_2$)] To freeze the root at zero temperature, \textit{and} ensure that the entropy difference between the root states is zero, we need $N_{02} = 0$. If $N_{12}=0$ as well, then the defect density ends up as $d_0=0$ and we are back to codeword-polarized BCs. Otherwise, any combination of $(N_{12}\geq 1, N_{22}=b-N_{12})$ is allowed. We find that the critical temperature increases as a function of $N_{22}$.
\end{description}

This leaves over 1000 possibilities, so we simplify matters by considering a few parameterized families.

First, we consider
\begin{equation}
\mathbf{N} = \begin{pmatrix}
N_{00} & 3 & 0 \\
0 & 0 & N_{12} \\
6-N_{00} & 3 & 6-N_{12}
\end{pmatrix} \Rightarrow \vec{d} = \frac{1}{9(4+N_{12}) - N_{00}(6 + N_{12})} \begin{pmatrix} 3 N_{12} \\ N_{12}(6-N_{00}) \\ 6(6-N_{00}) \end{pmatrix}.
\end{equation}

For all such families, the fraction of defects pushed off the root at $T=0$ is $5/12$. Therefore, the zero-temperature energy density is highest for the family with the largest $d_0$, which is $\mathbf{N}_{\rm max}$ above. In fact, for a fixed $N_{12}$ and varying $N_{00}$, $\varepsilon(x)/d_0$ follows nearly the same trend up to $x_c$. Note that $x_c$ does not depend on $N_{00}$.

We therefore now restrict our attention to $N_{00}=2$ (for which $d_0$ is largest) and vary the composition of the second column, keeping $N_{11}=0$:
\begin{equation}\label{eq:N-family}
\mathbf{N} = \begin{pmatrix}
2 & N_{01} & 0 \\
0 & 0 & N_{12} \\
4 & 6-N_{01} & 6-N_{12}
\end{pmatrix} \Rightarrow \vec{d} = \frac{1}{24 + (4+N_{01}) N_{12}} \begin{pmatrix} N_{01} N_{12} \\ 4 N_{12} \\ 24 \end{pmatrix}.
\end{equation}

\begin{figure}[hbtp]
\centering
\includegraphics[width=0.55\linewidth]{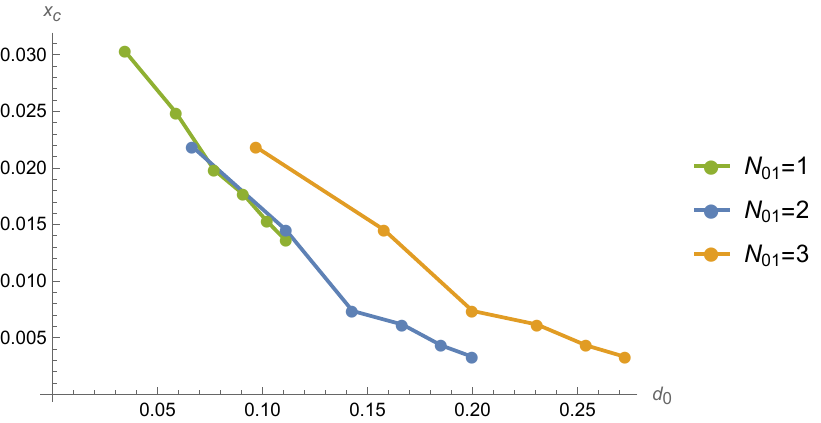}
\caption{Estimated $x_c$ vs. the density $d_0$ of type 0 branches, for three families of transfer matrices of the form~\autoref{eq:N-family}: $N_{01}=1$ (green), $N_{01}=2$ (blue), $N_{01}=3$ (orange), and varying $N_{12}$. For concreteness, since $x_c$ does not depend on the first column of $\mathbf{N}$, we take $N_{00}=2, N_{20}=4$ when computing $d_0$. The orange family includes the highest energy valley (\autoref{eq:N-max}).\label{fig:ferro-ish}}
\end{figure}

Two trends emerge, which lead to the conjectured region of thermally stable valleys as shown in~\autoref{fig:e-together}. First, for a given $N_{01}$, $x_c(\mathbf{N})$ decreases with increasing $N_{12}$ (\autoref{fig:ferro-ish}). This is because increasing $N_{12}$ makes the type 2 nodes less stable, while reducing the defect density. The green, blue, and orange curves in~\autoref{fig:ferro-ish} correspond to families with $N_{01} = 1, 2, 3$, respectively.

Second, for a given $N_{12}$, the defect density (as well as the energy density at a given temperature) increases with increasing $N_{01}$. A peculiarity of the Hamming [7,4,3] code is that the recursion relations for the type 1 root are identical for $N_{01}=2$ and $N_{01}=3$. Thus, the critical temperatures for $N_{01}=2$ and $N_{01}=3$ are the same (as can be seen in~\autoref{fig:ferro-ish}), and since $N_{01}=2$ leads to lower energy densities, it does not provide any additional bounds on the phase diagram. Thus, only the $N_{01}=1$ (green) and $N_{01}=3$ (orange) families are shown in~\autoref{fig:e-together}.

\subsection{Bulk energy density}\label{app:e-v2}
To determine the bulk energy density for these fine-tuned families, we proceed as follows. First, solve the coupled recursion relations for ($m_1,m_2$). If we join $b$ branches, with branch types $\vec{c}$, then conditioned on the root spin being $\sigma_0$, the root node has energy
\begin{equation}\label{eq:e-root}
\langle E(\spin_0,\vec{c}) \rangle = \frac{\sum_{\vec{\spin}} e^{-\beta E_v(\spin_0,\vec{\spin})} E_v(\spin_0,\vec{\spin}) \prod_{i=1}^b \left(\frac{(1 + m_{c_i} \spin_i}{2}\right)}{\sum_{\vec{\spin}} e^{-\beta E_v(\spin_0,\vec{\spin})} \prod_{i=1}^b \left(\frac{(1 + m_{c_i} \spin_i}{2}\right)}
\end{equation}
We then follow the average energy of nodes $r$ layers up from the root, starting from $r=0$. The update rule for the root type $\alpha$ is
\begin{equation}\label{eq:e-node-r}
\langle E(\spin_0,\vec{c}_\alpha)^{(r)} \rangle = \frac{1}{b} \frac{\sum_{\vec{\spin}} e^{-\beta E_v(\spin_0,\vec{\spin})} \prod_{i=1}^b \left(\frac{(1 + m_{c_{\alpha,i}} \spin_i}{2}\right)\sum_{i=1}^b \langle E(\spin_i,\vec{c}_{c_{\alpha,i}})^{(r-1)} \rangle}{\sum_{\vec{\spin}} e^{-\beta E_v(\spin_0,\vec{\spin})} \prod_{i=1}^b \left(\frac{(1 + m_{c_{\alpha,i}} \spin_i}{2}\right)}.
\end{equation}
As $r\rightarrow\infty$, this quantity becomes independent of the root type and the sign of the root spin, converging to the bulk energy density.

\section{Valley tracking at $\alpha=1$}\label{app:alpha1}
To make contact with the landscape realizable on closed graphs (as opposed to trees with fine-tuned boundaries), we now return to the $(\gamma,\alpha)$ ensembles, specializing to the line $\alpha=1$, $\gamma\beta=\beta_0$. With $\beta_0$ fixed, this choice of boundary conditions tracks the set of valleys that are typical at inverse temperature $\beta_0$. In this section, we provide some justification and intuition for why these BCs follow a set of valleys, solve analytically for $\ec{min}$ for a given $\beta_0$, and elaborate upon the interpretation of canonical and microcanonical instabilities. 

\subsection{Notation and sanity checks}
The $\alpha=1,\gamma\beta=\beta_0$ ensemble was introduced in Ref.~\cite{Krzakala2010,Zdeborova2010}, where it was presented as a method to adiabatically follow valleys away from the temperature where they are typical. We can translate their method into the language of this paper as follows. Choose some energy density $\ec{0}$. The temperature at which the paramagnetic energy density is $\varepsilon_0$ is the solution to  $\varepsilon_{\rm para}(x_0) = \varepsilon_0$. Now fix $x^{\gamma} = x_0$ and vary $\gamma$ (which also means varying $\beta$). Let $\varepsilon(x_0,x)$ denote the energy density in this valley at inverse temperature $\beta=-\log(x)$. 

This curve crosses the paramagnetic energy curve at $x=x_0$, i.e. $\gamma=1$. We will see below that indeed the only stable curves of constant $x_0$ do eventually intersect the paramagnetic energy curve, and thus are typical valleys at some temperature (namely, at inverse temperature $-\log(x_0)$).

Why should constant $x_0$ correspond to a tracking a particular set of valleys? Recall that in the $(\gamma,\alpha)$ parameterization, there are three relevant inverse temperatures: (1) the inverse temperature of the broadcast ($\gamma\beta/\alpha$), (2) the inverse temperature of the ``reweight'' at the boundary ($\alpha\beta$), and (3) the inverse temperature of the recursion from boundary towards the bulk ($\beta$). Specializing to $\alpha=1$, we can loosely think of the broadcast step as selecting the valley (i.e., the defect pattern) and the recursion step as setting the temperature \textit{within} that valley. Indeed, note that for $\alpha=1$, our leading-order expression for the defect density (\autoref{eq:defect-density}) evaluates to $\tilde{z}=2x_0(1+9x_0)$, with no dependence on $x$.

To make this argument more precise, let us perform two basic sanity checks. First, a few lines of algebra, using~\autoref{eq:beta-f},~\autoref{eq:Z-alpha1}, and~\autoref{eq:e-two-parts}, will convince the reader that  $\partial (\beta f(x_0,x))/\partial \beta = \varepsilon(x_0,x)$, as it must if we are populating the same valleys.

Second, at leading order, the configurational entropy density of the valleys being tracked by these BCs is constant. We can verify this by matching $f(x,x_0)$ onto $f(x,\gamma,\gamma)$ at each $x$ and applying~\autoref{eq:s-conf-gamma}. At order to which we have carried the expansion, the free energy density is a function only of $x$ and $\tilde{z}$, so matching the free energies at a given $x$ just means matching the defect densities.  Thus~\autoref{eq:s-conf-gamma} becomes
\begin{equation}\label{eq:s-conf-x0}
    s_{\rm conf}(x_0,x) = \log(2)/2 + 7 x_0 (1-\log(2x_0)) + O(x_0^2,x_0x).
\end{equation}

\autoref{eq:s-conf-x0} is sensible when we compare it to~\autoref{eq:s-conf-x} in the main text: all we have done is replace $x$ with $x_0$. This is closely related to the concept of using ``isocomplexity'' to define the bottoms of valleys (see Ref.~\cite{montanari2004cooling_schedule}).

\subsection{Zero temperature limit}
In the limit $x\rightarrow 0$ at fixed $x_0$, we obtain the energy density at the bottom of the chosen valley, $\ec{min}(\ec{0})\equiv\varepsilon(x_0,0)$. Leveraging an insight from Ref.~\cite{Zdeborova2010}, bulk observables can be determined analytically by expressing them in terms of the \textit{conditional} distribution $\tQ{\spin}{}$. 

Rather than considering the probability distribution of the root magnetization, we will track the energy difference at the root. That is, let $e\in\{-1,0,1\}$ denote the energy difference between the ground state with the root frozen to $+1$ and the root frozen to $-1$. Consider the distribution $\tilde{Q}_{\pm 1,\beta,1,\gamma}(e)$. In what follows, we will suppress the $(\beta,1,\gamma)$ subscript. 

As $x\rightarrow 0$, this distribution converges to
\begin{equation}\tilde{Q}_{\pm 1}^*(e) = c_{-1} \delta(e \pm 1) + c_0 \delta(e) + c_1 \delta(e \mp 1),  
\end{equation}
where $c_{-1} + c_0 + c_1 = 1$.

We derive a recursion relation for the coefficients $c_{\pm 1}$. These are closely related to the ``warning'' messages in the zero-temperature limit of belief propagation, called warning propagation~\cite{mezard2009information,Zdeborova2010}. For the Tanner-Hamming code, the resulting recursion relations are not terribly illuminating; the important takeaway is that the equations admit solutions only for $x_0\leq 0.00297...$, corresponding to $\varepsilon_0^*=0.0236...,T_0^*=0.1718...$. As noted in the main text, $T_0^*$ is slightly below the spin glass transition temperature $\Tglass = 0.176(1)$ numerically determined in Ref.~\cite{Placke_glass}. 

We split the energy density at $T=0$ into two parts, \'a la~\autoref{eq:e-two-parts}. The reason for splitting it this way is because if we instead took the $\beta\rightarrow\infty$ limit of~\autoref{eq:Ev-ga}, we would have to also keep track of the entropy difference on the root, which need not be restricted to a few values.

Explicitly, consider joining together $b+1$ branches at a common interaction vertex. We first draw a local configuration $(\spin_0,\vec{\spin})$ according to the Boltzmann weight $x_0^{E_v(\spin_0, \vec{\spin})}$. Then, we sample $e_i$ from the conditional distribution $\tilde{Q}_{\sigma_i}$ on the $i$th branch. If the vector $(e_0,\vec{e})$ is consistent with a codeword -- namely, if there exists a spin configuration such that every root is in its ground state and the new interaction node is also satisfied -- then $\ec{delta}(e_0,\vec{e}) = 0$. If not (for example if $e_0=+1$ and all other branches have $e_i=-1$), then $\ec{int}(e_0,\vec{e}) = 1$. For the Tanner-Hamming code, this evaluates to
\begin{equation}
    \ec{int}=\frac{7}{1 + 7 x} \left[x_0 (c_1^4 + 6 c_1^2 c_{-1}^2 + c_{-1}^4) + (1 + 3 x_0) c_1 c_{-1} (c_1^2 + c_{-1}^2)\right] \left[(1-c_0)^3 + 6 (1-c_0)^2 c_0 + 12 (1-c_0) c_0^2 + 4 c_0^3\right].
\end{equation}
$\ec{int}$ captures the change in the total energy when we introduce a new node. However, it is \textit{not} the local energy $\langle E_v \rangle$ on that node, because we might choose to flip one of the root spins from its ground state in order to satisfy the new node. Calculating that directly would require us to know the entropy gain from flipping one of the root spins.

Instead, we can appeal to \autoref{eq:Y-bulk} and subtract off the contribution from joining two branches at the root. Expressed in terms of $\tilde{Q}$, we draw $e_0, e_1$ from the \textit{same} conditional distribution, say $\tilde{Q}_{+1}$. A defect is created if $e_0=\pm 1, e_1 = \mp 1$: the two roots are frozen in opposite directions in their ground states. Therefore
\begin{equation}
    \ec{delta} = 2 c_1 c_{-1}.
\end{equation}
Altogether, therefore,
\begin{equation}
    \varepsilon(x_0,0) = \ec{int}(\vec{c}(x_0)) - \frac{b+1}{2} \ec{delta}(\vec{c}(x_0)).
\end{equation}
The difference $\ec{0} - \ec{min}(\ec{0})$ isolates the contribution of thermal excitations at the temperature at which the valley is typical, and is therefore an increasing function of $\ec{0}$.

\subsection{Canonical instability}
\begin{figure}[t]
\centering
\includegraphics[width=\linewidth]{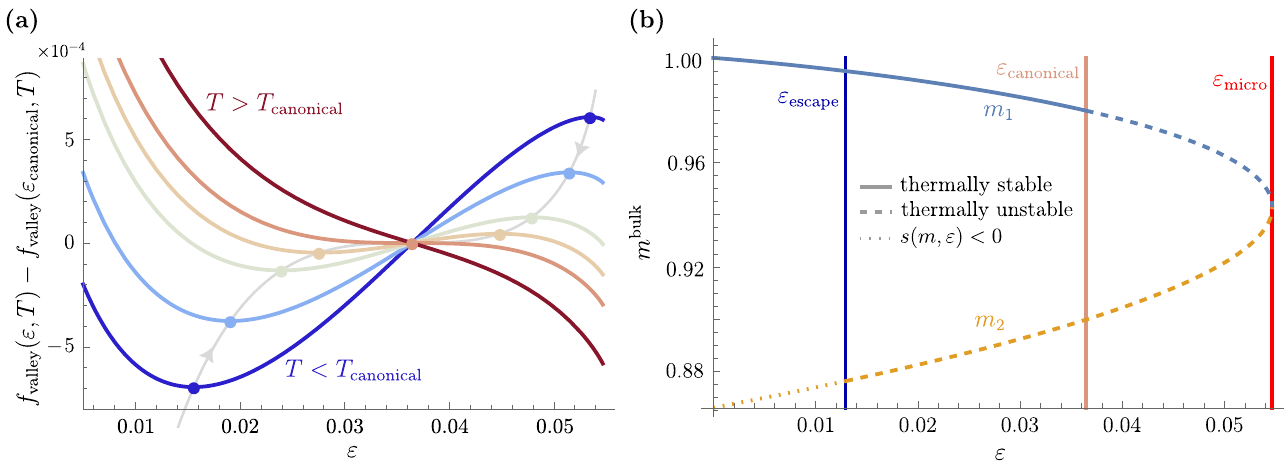}
\caption{(a) Mechanism for the canonical instability within a valley, illustrated for ground state valleys. $f_{\rm valley}$ is the free energy density, with the vertical axis defined so that all curves intersect at $(\ec{canonical},0)$. As temperature increases (indicated by the direction of the gray arrows), the two extrema (marked points) approach each other, colliding at an inflection point. For $T>T_{\rm canonical}$, there is no local minimum within the valley.\label{fig:canonical-instability} (b) Bulk magnetization vs. energy density at the ferromagnetic fixed points. The top half of the curve (blue) is $m_1$, where $s(m,\varepsilon)$ has a local maximum associated with states within the valley. The bottom half of the curve (orange) is $m_2$, where $s(m,\varepsilon)$ has a local minimum associated with the saddle to exit the valley. Solid and dashed segments indicate thermally stable and unstable fixed points (i.e., local minima and maxima of the bulk free energy at the corresponding temperature); the dotted portion of the lower curve is both thermally unstable on the tree and vanishingly rare on the closed graph ($s(m_2,\varepsilon)<0$). Vertical lines from left to right denote $\ec{escape}, \ec{canonical}, \ec{micro}$.}
\end{figure}

The mechanism for the canonical instability is sketched in~\autoref{fig:canonical-instability}a, using free energy curves for the valleys containing ground states.

Recall from the main text that at low temperatures, the within-valley free energy density $f_{\rm valley}(\varepsilon,T)$ takes the qualitative form of the blue curves in~\autoref{fig:canonical-instability}a: it has a local minimum within-valley energy density $\ec{valley}(T)$, and a local maximum at a higher energy density $\ec{barrier} > \ec{valley}$: the barrier between being trapped in the valley and escaping to the paramagnet. For a set of valleys that are typical at temperature $T_0$, we can identify $\ec{valley}(T)$ and $\ec{barrier}(T)$ as, respectively, the bulk energy densities $\varepsilon(x_0,x)$ and $\ec{unstable}(x_0,x)$ evaluated at stable and unstable fixed points of the recursion. The difference in bulk free energy densities evaluated at the respective fixed points
\begin{equation}
    \Delta f_{\rm valley}(T) = f(x_0,x) - f_{\rm unstable}(x_0,x)
\end{equation}
quantifies the bottleneck for local dynamics.

As $T_{\rm canonical}$ is approached, $\ec{valley}(T)$ increases, $\ec{barrier}(T)$ decreases, and the height of the barrier $\Delta f_{\rm valley}(T)$ decreases, until at $T=T_{\rm canonical}$, the two extrema collide at an inflection point $\ec{canonical}$. 

\subsection{Microcanonical instability}

Turning to the microcanonical ensemble, we again focus on the valleys containing ground states for concreteness and to enable exact calculations; the generalization to higher-energy valleys is straightforward. At a given $T<\Tdyn=T_{\rm canonical}(0)$ (a vertical slice through~\autoref{fig:emin-escape}b), there exist two nontrivial ferromagnetic fixed points: a stable fixed point at magnetization $m_{\rm ferro}$ and energy density $\ec{ferro}$, and an unstable one at magnetization $m_{\rm unstable}$ and energy density $\ec{unstable}$. Here, the designation of ``stable'' vs. ``unstable'' is implicitly within the canonical ensemble, at a given temperature $T$. 

Now suppose we instead take a horizontal slice through~\autoref{fig:emin-escape}b, at fixed $\varepsilon$. For $\varepsilon \leq \ec{canonical}$, this horizontal cut intersects the thermally stable energy curve $\ec{ferro}(T)$ at temperature $T_1$, and the thermally unstable energy curve $\ec{unstable}(T)$ at $T_2 < T_1$. The corresponding bulk magnetizations are $m_1 = m^{\rm bulk}_{\rm ferro} > m_2 = m^{\rm bulk}_{\rm unstable}$.\footnote{In the ferromagnetic valleys, \autoref{eq:mbulk} reduces to $m^{\rm bulk}_* = \frac{2m_*}{1+m_*^2}$ where $m_*$ is the root magnetization at a fixed point. Henceforth we will drop the superscript ``bulk'' and will use $m$ to denote a bulk magnetization, as bulk quantities are what we directly measure on closed graphs.} The fixed point at $m_1$ is \textit{both} a local minimum of the within-valley free energy as a function of $\varepsilon$, and a local maximum of the entropy density in the microcanonical ensemble $s(m,\varepsilon)$. The fixed point at $m_2$ is both a local maximum of $f_{\rm valley}$ (the saddle point to exit the valley in the canonical ensemble) and a local minimum of $s(m,\varepsilon)$ (the saddle point to exit the valley while remaining at energy density $\varepsilon$, provided $s(m_2,\varepsilon) \geq 0$). These two fixed points are the solid blue and dashed orange curves in~\autoref{fig:canonical-instability}b.

We can continue to follow $m_1,m_2$ as $\varepsilon$ is increased beyond $\ec{canonical}$, where now both fixed points lie on the thermally unstable branch. That is, although the thermally stable branch no longer exists for $\varepsilon > \ec{canonical}$, $\ec{unstable}(T) = \varepsilon$ has two solutions (dashed blue and dashed orange curves), which are still local extrema of $s(m,\varepsilon)$. The solutions collide and annihilate at $\varepsilon=\ec{micro}$, beyond which there is no well-defined valley population. 

Finally, let us elaborate on how the curves in~\autoref{fig:emin-escape}c and \autoref{fig:canonical-instability}a were obtained.

The free energy curve $f_{\rm valley}(\varepsilon,T)$ for $\varepsilon\leq \ec{micro}$ is reconstructed from
\begin{equation}
    f_{\rm valley}(\varepsilon, T)= \varepsilon - T s(m_1(\varepsilon), \varepsilon).
\end{equation}
Note that, as its name suggests, $f_{\rm valley}$ only depends on information from the ``within-valley'' fixed points $m_1$, not the saddle point at $m_2$.

To obtain the full $s(m,\varepsilon)$ curve away from its extrema, we introduce a bulk external field. At a given temperature, the external field $h$ modifies the fixed point magnetizations, and the bulk free energy density is defined as~\cite{Bellitti2021}
\begin{equation}
    f(m,T) = f(h,T) - \frac{b+1}{2} m h.
\end{equation}
The corresponding energy density is expressed in the usual way in terms of messages received by an interaction node (\autoref{eq:Ev-ga}). Let $T(m,\varepsilon)$ denote the unique solution to the equation $\varepsilon(m,T) = \varepsilon$.  Then
\begin{equation}
    s(m,\varepsilon) = \frac{1}{T(m,\varepsilon)} \left[\varepsilon - f(m,T(m,\varepsilon))\right].
\end{equation}

\section{Flat histogram techniques}\label{app:multi}
In this section we provide more detail on the numerical exploration of the model on closed graphs. We describe the adaptive histogram method for both $x(E)$ and $x(M)$ and present additional numerical evidence for the phase diagram sketched in~\autoref{fig:emin-escape}.

\subsection{Finding the bottom of the valley}\label{app:initialize}

In the main text, we skirted the issue of initialization: how do we start in a valley whose minimum energy density is $\ec{min}$?

One way to do this is to sample a state from the global Boltzmann distribution with energy $E_0$, then search for the bottom of the well containing that initial state, $E_{\rm min}$, by running adaptive Metropolis with $x(E)$ (detailed in the next subsection) biased towards lower energies. This allows us to relax towards the bottom of the well without getting stuck in shallow local minima (i.e., ``ridges'' within the valley).\footnote{Empirically, a quench to zero temperature often reaches the same minimum, though it can fail for higher-energy valleys.} A given $E_0$ induces a distribution over $E_{\rm min} \leq E_0$, peaked at some $\overline E_{\rm min}(E_0)$. As $E_0$ increases, the distribution broadens, and $E_0 - \overline{E}_{\rm min}(E_0)$ increases due to the increasing contribution of thermal excitations. Note that the tree calculations in the previous section rest on the reasonable physical assumption that the distribution of $E_{\rm min}/N_v$ is sharply peaked as $N_v\rightarrow \infty$, allowing us to uniquely define $\ec{min}$ as a function of $\ec{0}$ in that limit. 

If $\varepsilon_0>\Eglass$, a typical initial state will not be trapped in a stable valley, and thus we cannot use this approach to probe valleys above $\ec{min} = \overline{\varepsilon}_{\rm min}(\Eglass)$; stable valleys may exist at higher energy densities, but they will be atypical within the global Gibbs state.

Consider a valley whose bottom is at energy $E_{\rm min}$, and let $\Sigma_{\rm min}$ denote the set of all states inside the valley at that energy. As $E_{\rm min}$ increases, the size of this set increases: the valley becomes more like a shallow bowl than a deep, narrow well. A rough estimate for the degeneracy comes from the within-valley entropy density for $\gamma=\alpha=1$ at energy $\varepsilon$.\footnote{More precisely: $\exp(N_v s(\varepsilon_0,1,1))$ counts the states within a typical valley at energy density $\varepsilon_0$. The number of states at the bottom of this valley is slightly lower.} Let $\vec{\spin}_{\rm min}$ denote the average over the configurations in the set $\Sigma_{\rm min}$.

During the initialization, we record the unique spin states at the lowest energy visited so far. If we find the true bottom of the valley and sufficiently explore it, then at the end of the initialization step we have visited a representative subset of $\Sigma_{\rm min}$. We average over this subset to obtain an estimate of $\vec{\spin}_{\rm min}$. The correlation 
\begin{equation}\label{eq:M}
    M(\vec{\sigma}) = \vec{\sigma} \cdot \vec{\spin}_{\rm min}
\end{equation}
serves as a rough measure of the ``distance'' of a state $\vec{\sigma}$ from the bottom of the valley.\footnote{Note that when exploring valleys containing ground states, we can focus without loss of generality on the all-up ground state. Then, $M(\vec{\spin})$ is simply the net magnetization $\sum_i \spin_i$, hence the choice of letter.}

\subsection{Canonical instability: $x(E)$}

Consider local Metropolis dynamics at inverse temperature $\beta$. If the dynamics is ergodic within the full configuration space, then the steady-state energy histogram is 
\begin{equation}\label{eq:n-x0}
    n(x, E) \propto x^E e^{S(E)} = e^{-\beta F(E,T)}.
\end{equation}
Here, $F(E,T)$ is the ``global free energy'', which in the absence of redundancies can be computed exactly, since $S(E) = K\log 2 + E \log(2^{\nL-kL}-1) + \log\binom{N_v}{E}$. As it must, this free energy has a unique global minimum at $E=\ec{para}(x)$ [\autoref{eq:e-para}].

However, at low temperatures and finite time, due to extensive free energy barriers, a system initialized within a valley will remain trapped within that valley, and $F(E,T)$ in~\autoref{eq:n-x0} is replaced by the within-valley free energy $F_{\rm valley}(E,T)$. 

\subsubsection{Multicanonical method}
When trying to quantify the height of the barrier and the temperature at which it vanishes, standard single-temperature Metropolis encounters a problem: the states at the top of the free energy barrier are exponentially rare in $\Delta F$. The multicanonical method~\cite{Berg1991,Berg1992,Janke1998,Mau1999} presents a workaround. Rather than accepting proposed spin flips according to a single Boltzmann factor $x$, we define an energy-dependent Boltzmann factor $x(E)$, such that a proposed flip that would change the energy from $E$ to $E+\Delta E$ is accepted with the probability given in~\autoref{eq:P-accept}. In the standard implementation of the method, we seek $x(E)$ such that $n[x(E),E]$ is independent of $E$ within some prescribed interval. By definition, if we had already determined $S_{\rm valley}(E)$, we could simply read off
\begin{equation}\label{eq:xE-target}
    x(E) = e^{S_{\rm valley}(E+1) - S_{\rm valley}(E)}.
\end{equation}
Note that $x(E)$ defines a ``microcanonical'' within-valley temperature, in the sense that $x(E) = e^{-\beta_{\rm valley}(E)}$. 

If we first run the dynamics at a single temperature, our initial guess for $x(E)$ will therefore be
\begin{equation}
x(E) = x \frac{n(x,E)}{n(x,E+1)}.
\end{equation}
In practice, as we have emphasized, the fixed-temperature run will not provide sufficient precision near the barrier. Thus, our task will be to iteratively update $x(E)$ until the flatness criterion is satisfied. Given an energy histogram obtained by a large number of sweeps using $x(E)$, we then update $x(E)\rightarrow x_{\rm new}(E)$ where
\begin{equation}\label{eq:n-update}
    x_{\rm new}(E) = \frac{n(E)}{n(E+1)} x(E).
\end{equation}

The full algorithm is presented in~\autoref{alg:adaptive-energy-window}. It takes as input the initial state $\vec{\sigma}$, obtained from the initialization procedure discussed in~\autoref{app:initialize} above, along with an initial guess for the adaptive Boltzmann factor $x(E)$; an initial interval $[E_1,E_2]$ on which $x(E)$ satisfying~\autoref{eq:xE-target} is to be learned; the amount $E_{\rm step}$ by which the upper limit of this interval will be incremented after a successful iteration; and several additional parameters. We generally set $E_1$ equal to $E_{\rm min}$, so that failure to equilibrate the histogram down to $E_{\rm min}$ signals a potential escape from the valley. This also allows us to monitor whether a lower energy state is ever discovered, which would indicate that our initialization step did not sufficiently explore the valley. This occurs in rare instances for $\ec{0} \gtrsim 0.02$. In the special case $E_{\rm min}=0$, since we know for certain that no lower energy states can be found, we choose $E_1$ to be larger and also set a minimum energy cutoff to avoid wasting time exploring near the bottom of the valley (which is far below the instability energy). 

To learn $x(E)$ on the interval $[E_1, E_2]$, we track the histogram of energies visited on that interval  during each round, where round $i$ consists of $N_{\rm sweeps}(i)$ Monte Carlo sweeps (lines 5-9).\footnote{We take $N_{\rm sweeps}(i)$ to be an increasing function of $i$, so that more data are collected as the energy histogram becomes flatter.} Outside the learning interval, $x(E)$ is taken to be a small constant, i.e. the dynamics outside $[E_1, E_2]$ is standard Metropolis at low temperature, which as $E_2$ is increased means that, in practice, the majority of the counts will fall in the interval $[E_1, E_2]$. After each round, $x(E)$ within the interval is updated using an optimized version of~\autoref{eq:n-update}.\footnote{\label{foot:opt}Some optimizations include: only updating $x(E)$ for bins that have a sufficient number of counts; aggregating histogram data for energy bin $E$ over all previous rounds where $x(E), x(E+1)$ were not updated; setting a maximum on $x(E)$ to avoid getting led astray in the early rounds of learning; and smoothing out $x(E)$ by averaging over a sliding window.} If round $i$ satisfies a predetermined criterion --- e.g., the variance of the energy distribution on $[E_1,E_2]$ is below some threshold, the minimum number of counts exceeds some threshold, and $i$ exceeds a minimum number of rounds---then we consider $x(E)$ to have been learned sufficiently well on that interval, save $(\vec{\spin}, E_2)$ as the most recent successful iteration, and extend the learning interval to $[E_1,E_2 + \Delta E]$ (lines 10-14,19). If the flatness criterion is not met after $N_{\rm rounds}$ rounds, which can occur for higher-energy valleys where the energy landscape is less smooth and the dynamics is ``glassier,'' then we revert to the most recent successful iteration, reduce the size of the energy increment, and try again on a smaller learning interval (lines 16-17). The entire algorithm terminates in success when $E_2$ exceeds a preselected value, or terminates in failure when $\Delta E < E_{\rm min-step}$ (lines 17-18, 20).

\begin{algorithm}[t]
\caption{Adaptive energy-window Monte Carlo sampling}
\label{alg:adaptive-energy-window}
\LinesNumbered
\KwIn{\begin{tabular}[t]{@{}l@{}}
    \textsc{Initial conditions:}  state $\vec{\spin}$; energy window $[E_1, E_2]$; energy increment $E_{\rm step}$; adaptive Boltzmann factor $x(E)$ \\
    \textsc{Parameters:} Allowed configuration space $\Sigma$; maximum number of rounds $N_{\rm rounds}$;
    number of sweeps in $i$th round \\ $N_{\rm sweep}(i)$;
    minimum energy increment $E_{\rm min-step}$; target energy $E_{\rm target}$;
    histogram acceptance function $\texttt{criterion}$
    \end{tabular}
    }
\KwOut{$x(E)$}

Set $\Delta E, \vec{\spin}_{\rm success}, E_{\rm success} \gets E_{\rm step}, \vec{\sigma}, E_2$\;

\While{$E_2 \leq E_{\rm target}$}{
    $\texttt{roundSuccess} \gets \mathrm{false}$\;

    \For{$i \gets 1$ \KwTo $N_{\rm rounds}$}{
        Initialize an empty energy histogram $H_i(E)$ on $[E_1,E_2]$\;

        \For{$s \gets 1$ \KwTo $N_{\rm sweep}(i)$}{
            $\vec{\spin} \gets$ \MCSweep{$\vec{\spin}, x(E), \Sigma$}\; 

            Update $H_i(E)$ with state $\vec{\spin}$\;
        }

        Update $x(E)$ on $[E_1,E_2]$ using the chosen variant of
        \autoref{eq:n-update}\;

        \If{$\mathtt{criterion}(H_i,i)$}{
            $\mathtt{roundSuccess} \gets \mathrm{true}$\;
            \textbf{break}\;
        }
    }

    \eIf{$\mathtt{roundSuccess}$}{
        $\vec{\sigma}_{\rm success}, E_{\rm success} \gets \vec{\sigma}, E_2$\;
    }{\tcp{Revert to most recent successful state}
        $\Delta E,\vec{\sigma}, E_2, \gets \Delta E/2, \vec{\sigma}_{\rm success}, E_{\rm success}$\;

        \If{$\Delta E<E_{\rm min-step}$}{
            \Return{$x(E)$}\;
        }
    }
    $E_2 \gets E_2 + \Delta E$\;
}

\Return{$x(E)$}\;
\vspace{10pt}
\Fn{\MCSweep{$\vec{\spin},x(E),\Sigma$}}{\nllabel{line:run-round}
\For{$j \gets 1$ \KwTo $\nL N_v/2$}{
$\vec{\spin}' \gets$ flip spin $j$ of configuration $\vec{\spin}$\;
\If{$\vec{\spin}' \in \Sigma$}{Set $\vec{\spin} \gets \vec{\spin}'$ with probability $P_{\rm accept}$ [\autoref{eq:P-accept}]\;}}
    \KwRet{$\vec{\spin}$}\;
}
\end{algorithm}

\subsubsection{Space of allowed configurations}
Let us elaborate on the space of allowed configurations $\Sigma$, which is an input parameter to~\autoref{alg:adaptive-energy-window} and its subroutine, \MCSweep{} (\autoref{line:run-round}). To properly sample only the within-valley free energy and entropy, we need a way to prevent the state from escaping the valley. 

One approach, used to produce~\autoref{fig:emin-escape}b and~\autoref{fig:canonical-instability}, is to reject all proposed configurations for which $M(\vec{\sigma})<M_{\rm min}$, where $M(\vec{\spin})$ is the correlation defined in~\autoref{eq:M}. This ensures that we do not sample states that are too far away from the bottom of the valley. The choice of $M_{\rm min}$ requires some finesse: too high of a cutoff will exclude high-energy, low-$M$ states within the valley, but if the cutoff is too low, then the system can escape. In practice, we observe the latter: as $E_2$ becomes large, in attempting to equilibrate to a flat histogram on the interval $[E_1,E_2]$, we begin to populate states outside the valley. This can be detected as a secondary peak in the distribution of $M$ near $M_{\rm min}$, which results in an erroneous upturn in $x(E)$ at large $E$. Fortunately, these escapes generally only occur at energy scales beyond the canonical instability (they are closely tied to the vanishing of the entropy barrier at the microcanonical instability), so we can safely truncate the $x(E)$ curve before the spurious upturn. 

Another approach is to impose a cutoff on the maximum energy $E_{\rm max}$, i.e., reject all moves that result in a configuration with energy $E>E_{\rm max}$. For example, we could set $E_{\rm max}$ equal to $E_2$. Then, as $E_2$ is slowly increased, we not only learn the location of the canonical instability (by learning $x(E)$)  but also learn an upper bound on the ``escape energy''. That is, let $E^*_{\rm max}$ denote the lowest $E_{\rm max}$ for which the system escapes the valley, as signaled by a sharp drop in $M$ and subsequent failure to equilibrate the counts in the interval $[E_1, E_{\rm max}]$. Then $E^*_{\rm max}$ upper bounds the energy $E_{\rm escape}$ at which there exists a path to escape the valley. Note that $E_{\rm escape} \leq E_{\rm micro}$: at $E_{\rm escape}$, there exist \textit{some} paths to exit the valley, but the time scale to escape remains exponentially long in the height of the entropy barrier. In practice, the upper bound on $E_{\rm escape}$ obtained in this way is incredibly loose: for ferromagnetic valleys, the zero of $s(m_2,\varepsilon)$ implies $\ec{escape}\approx 0.0129$, but on closed graphs we do not observe escapes until $\varepsilon \approx 0.05$.

The danger of using $E_{\rm max}$ alone without a cutoff on $M$ is twofold. First, imposing a sharp upper limit on $E$ can cut off certain pathways between configurations within the valley (as when there are ``ridges'' within the energy landscape). Empirically, however, we find that as $E_{\rm max}$ is slowly increased, $x(E)$ remains stable and insensitive to the cutoff, suggesting that this effect is insignificant. The second problem is that escapes from the valley generally occur at lower energies than if we impose a lower limit on $M$, since only imposing a limit on $E$ does not forbid rare escape routes that appear above $E_{\rm escape}$. For this reason, this second method is generally inferior, though we emphasize that for the valleys and ranges of energies where we tried both methods, the results agree. 

\subsubsection{Generalizations}
Note that while the update criterion~\autoref{eq:n-update} aims for a flat histogram on the interval $[E_1,E_2]$, we can modify it to target any functional form (i.e. a target histogram $n_{\rm target}(E)$) by taking:
\begin{equation}
    x_{\rm new}(E) = \frac{n(x(E),E)}{n(x(E),E+1)} \frac{n_{\rm target}(E+1)}{n_{\rm target}(E)} x(E).
\end{equation}
In the initialization step, we take $n_{\rm target}(E)$ to be linearly decreasing in $E$, thus biasing towards states at the bottom of the well.

\subsection{Microcanonical instability: $x(M)$}

Now let $\Sigma(E,\Delta E)$ denote the set of all states within a microcanonical shell $[E-\Delta E, E+\Delta E]$. If we consider the intersection of that microcanonical shell with a chosen valley, then the density of states $n(E,M)$ is peaked at some $M_{\rm valley}^*(E)$. On the other hand, across the entire microcanonical shell, $n(E,M)$ has a global maximum at $M_{\rm para}^*(E)=0$. The microcanonical instability occurs when the entropy barrier (that is, a minimum in $n(E,M)$ as a function of $M$) between them vanishes.

Before proceeding, let us comment on the relation to the entropy barriers alluded to in the previous subsection. There, we sampled states across a range of energies starting in a chosen valley, tuning the transition probabilities such that the energy histogram is flat on some interval $[E_1, E_2]$. If we discard all states outside this interval\footnote{Recall that even when we do not impose an upper cutoff on $E$, the vast majority of counts do fall in the interval $[E_1, E_2]$.}, then the induced probability distribution on $M$ is the integral of the normalized density of states $ \propto n(E,M)$ over that energy interval. This integrated distribution also has a tail to smaller $M$. If the state escapes the valley, then the dynamics begins to explore the space of lower-$M$ states. The ideal choice of $M_{\rm min}$ is therefore one that lies ``inside'' the entropy barrier of the total integrated distribution: i.e., in the far left tail of the within-valley distribution (so that we are not forbidding too many within-valley states), and the far right tail of the distribution just outside the valley (so that, even if the system makes a ``partial'' escape from the valley, it will tend to return, due to the low entropy of allowed outside-valley states). As $E_2$ increases, the barrier shifts and eventually disappears, beyond which the valley ceases to be well-defined in terms of the coordinate $M$. Thus, in principle, we can learn $x(E)$ up to the energy of the microcanonical instability, though in practice, this would require optimizing $M_{\rm min}$ as $E_2$ is increases. 

A more direct and precise way of learning the microcanonical instability energy is to apply the flat histogram method to $M$ instead of $E$, in order to sample the states near the entropy barrier at a given energy.

The algorithm is as follows. First, we find the bottom of the valley as described in~\autoref{app:initialize}. Then, after heating up within the valley until reaching a state in the energy window $\Sigma(E,\Delta E)$, we perform a large number of Monte Carlo sweeps within that window, accepting a proposed move with probability 1 if it remains within the window and rejecting otherwise. During this long sweep, we track $M(\vec{\sigma})$ and identify the mode of its distribution, $M^*$. 

\begin{algorithm}[t]
\caption{Adaptive microcanonical Monte Carlo sampling}
\label{alg:adaptive-micro}
\LinesNumbered
\KwIn{\begin{tabular}[t]{@{}l@{}}
    \textsc{Initial conditions:}  state $\vec{\spin}$; magnetization window $[M_1, M_2]$; increment $M_{\rm step}$; adaptive Boltzmann factor $x(M)$ \\
    \textsc{Parameters:} Allowed configuration space $\Sigma$; maximum number of rounds $N_{\rm rounds}$;
    number of sweeps in $i$th round \\ $N_{\rm sweep}(i)$;
    minimum increment $M_{\rm min-step}$; target magnetization $M_{\rm target}$;
    histogram acceptance function $\texttt{criterion}$
    \end{tabular}
    }
\KwOut{$x(M)$}

Set $\Delta M, \vec{\spin}_{\rm success}, M_{\rm success} \gets M_{\rm step}, \vec{\sigma}, M_1$\;

\While{$M_1 \geq M_{\rm target}$}{
    $\texttt{roundSuccess} \gets \mathrm{false}$\;

    \For{$i \gets 1$ \KwTo $N_{\rm rounds}$}{
        Initialize an empty ``magnetization'' histogram $H_i(M)$ on $[M_1,M_2]$\;

        \For{$s \gets 1$ \KwTo $N_{\rm sweep}(i)$}{
            $\vec{\spin} \gets$ \MCSweepM{$\vec{\spin}, x(M), \Sigma$}\; 

            Update $H_i(M)$ with state $\vec{\spin}$\;
        }

        Update $x(M)$ on $[M_1,M_2]$ using the chosen variant of
        \autoref{eq:n-update}\;

        \If{$\mathtt{criterion}(H_i,i)$}{
            $\mathtt{roundSuccess} \gets \mathrm{true}$\;
            \textbf{break}\;
        }
    }

    \eIf{$\mathtt{roundSuccess}$}{
        $\vec{\sigma}_{\rm success}, M_{\rm success} \gets \vec{\sigma}, M_1$\;
    }{\tcp{Revert to most recent successful state}
        $\Delta M,\vec{\sigma}, M_1, \gets \Delta M/2, \vec{\sigma}_{\rm success}, M_{\rm success}$\;

        \If{$\Delta M<M_{\rm min-step}$}{
            \Return{$x(M)$}\;
        }
    }
    $M_1 \gets M_1 - \Delta M$\;
}

\Return{$x(M)$}\;
\vspace{10pt}
\Fn{\MCSweepM{$\vec{\spin},x(M),\Sigma$}}{\nllabel{line:run-round}
\For{$j \gets 1$ \KwTo $\nL N_v/2$}{
$\vec{\spin}' \gets$ flip spin $j$ of configuration $\vec{\spin}$\;
\If{$\vec{\spin}' \in \Sigma$}{Set $\vec{\spin} \gets \vec{\spin}'$ with probability $P_{\rm accept} = \mathrm{\autoref{eq:p-accept-m}}$}}
    \KwRet{$\vec{\spin}$}\;
}
\end{algorithm}

The rest of the algorithm, detailed in~\autoref{alg:adaptive-micro}, proceeds analogously to~\autoref{alg:adaptive-energy-window}, but with $E$ replaced by $M$. The function $x(M)$ is learned on an interval $[M_1, M_2]$, and $M_1$ is slowly reduced. We take $M_2=M^*(E)$, since we are not interested in learning the right tail of the within-valley distribution. The allowed configuration space $\Sigma$ is now the subset of states in the microcanonical shell with $M$ exceeding a chosen cutoff; we take this cutoff to be $M_{\rm min} = M_1$, thus expanding the configuration space as the learning interval is also extended. 

A proposed update from $\vec{\spin}$ to $\vec{\spin}'$ is rejected if $\vec{\spin}'\notin \Sigma$, and otherwise accepted with probability
\begin{equation}\label{eq:p-accept-m}
P_{\rm accept}(\vec{\sigma}\rightarrow \vec{\spin}') = \begin{cases}
\min\left(1, \left[x(M(\vec{\sigma}))\right]^{M(\vec{\sigma}) - M(\vec{\sigma}')}\right) & M(\vec{\sigma}') < M(\vec{\sigma}) \\
\min\left(1, \left[x(M(\vec{\sigma}'))\right]^{M(\vec{\sigma}) - M(\vec{\sigma}')}\right) & \mathrm{otherwise}.
\end{cases}
\end{equation}

Once $x(M)$ has been updated to make a flat histogram over $M$, it satisfies $x(M) = e^{\partial_M S(E,M)}$,
allowing us to read off $S(E,M)$ as approximately\footnote{In practice, we discretize $M$ with a spacing of 2 when defining and integrating over $x(M)$, since flipping a single spin changes the ``effective magnetization'' $M(\vec{\spin})$ by exactly 2 if the state at the bottom of the valley is unique.}
\begin{equation}\label{eq:m-barrier}
    S(M,E)-S(M^*(E),E) = - \int_{M}^{M^*} dM \log(x(M)).
\end{equation}
In particular, the local extrema of $S(M,E)$ occur at the values of $M$ for which $x(M)=1$. The entropy barrier corresponds to a local minimum of $S(M,E)$, at $M=M_{\rm barrier}(E)$.

\begin{figure}[t]
\centering
\includegraphics[width=0.45\linewidth]{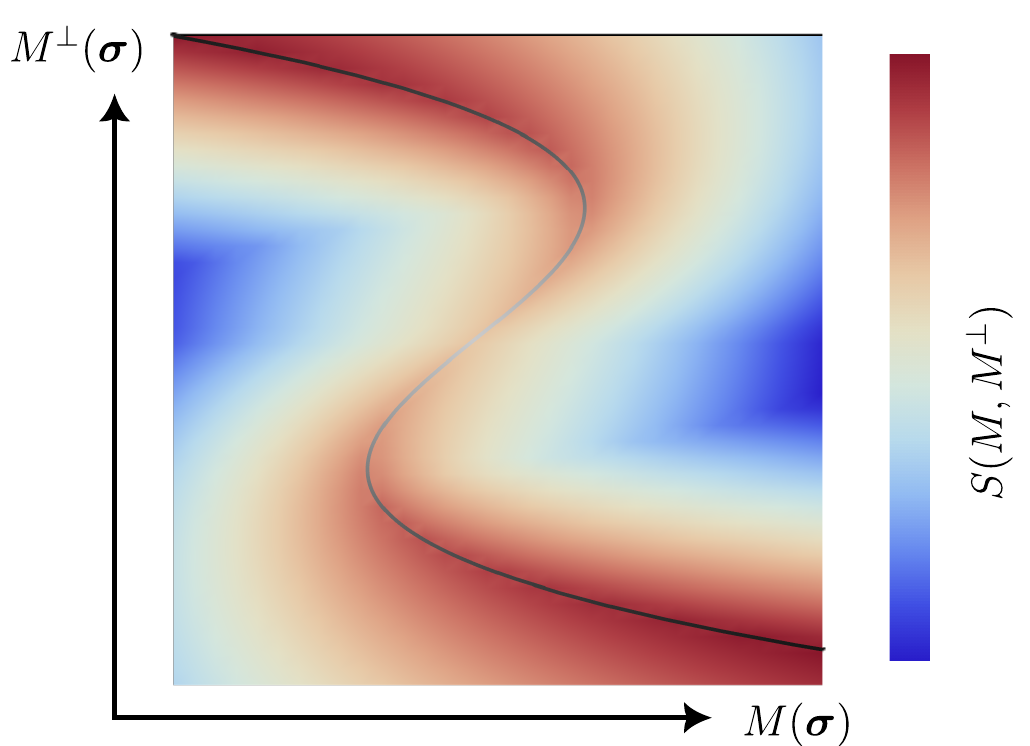}
\caption{\label{fig:entropy}Sketch of how configuration space might look when projected onto two coordinates, $M$ and $M^\perp$, instead of one. Red (blue) indicates a larger (smaller) density of states. The highest-entropy route from high $M$ to low $M$ is along the gray curve, but estimating the entropy in terms of $M$ alone will fail to capture the paucity of states near the inflection point of this curve.}
\end{figure}

In the End Matter, we referred to the difference $S(E,M^*(E))-S(E,M_{\rm barrier}(E))$ as a \textit{lower bound} on the true entropy barrier. This is because the actual entropy landscape at a given $E$ is defined over a high-dimensional configuration space. When we write $S$ as a function of a single additional coordinate $M(\vec{\sigma})$, we are integrating over the dimensions orthogonal to $M$, summing the contributions from states that may be separated by many local moves. A minimal example is sketched in~\autoref{fig:entropy}. If such a scenario occurs, then a better choice of reaction coordinate would be the distance along the gray curve.

With this caveat in mind, we compare the lower bounds on the entropy barrier around codeword valleys on system size $N_v=1298$ with the entropy barrier $\Delta S$ inferred from the exponential scaling of the escape time, $\tau(E) \sim \exp(\Delta S(E)/T)$. As a function of the microcanonical energy $E$, the escape time follows the rough trend $\log(\tau(E)) \approx (S(E,M^*(E))-S(E,M_{\rm barrier}(E))) + c$, suggesting that at least for these valleys, the lower bound qualitatively captures the true entropy barrier.

\subsection{Supporting numerics}
\subsubsection{Simulation details}
To perform simulations on closed graphs, we first use a version of the method described in Ref.~\cite{linial2019rhgrg} (see App. C of Ref.~\cite{Placke_glass} for details) to generate a high-girth random regular graph (HGRRG). The greedy algorithm succeeds with high probability if, for a target girth  $g$ (where the girth is the length of the shortest cycle), we take
\begin{equation}
N_v(g) = b^{g-1}+2 + [b \mathrm{\, mod \, } 2].
\end{equation}
Plugging in $b=6$ for the Tanner-Hamming code family, we consider the sequence of system sizes $N_v(4)=218, N_v(5)=1298, N_v(6)=7778$. All data presented were averaged over 20-100 different graph realizations. 

\begin{figure}[t]
\centering
\includegraphics[width=0.5\linewidth]{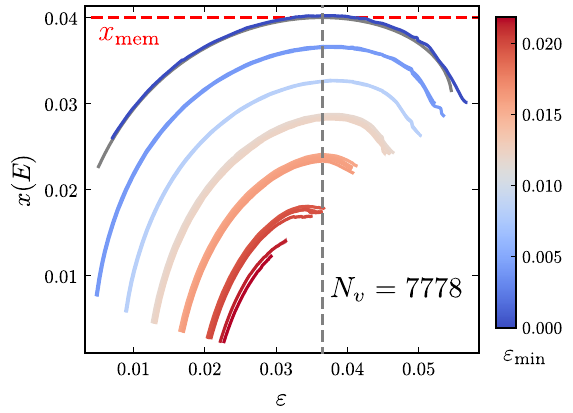}
\caption{$x(E)$, the adaptive Boltzmann factor needed to make a flat histogram, as a function of energy density $\varepsilon$, on closed graphs with $N_v=7778$ (girth 6). The gray curve is $x(E)$ for ferromagnetic valleys deduced from the tree calculation, i.e. $\exp(-\partial s/\partial \varepsilon)$.\label{fig:xE}}
\end{figure}

In the analysis of the canonical instabilities, we first examined $N_v=1298$ before scaling up to $N_v=7778$, observing a small downward shift in $x_{\rm canonical}(\ec{min}=0)$ from $\approx 0.041$ for $N_v=1298$ to $\approx 0.0402$ for $N_v=7778$. At the larger system size, $x_{\rm canonical}(0)$ is only slightly higher than the codeword-polarized tree calculation $x_{\rm mem}=1/25$, and likewise $\ec{canonical}$ across a wide range of $\ec{min}$ is very close to $\ec{ferro}(x_{\rm mem})$ obtained from a tree calculation, indicating that the finite-size/finite-girth effects at $N_v=7778$ are already quite mild (\autoref{fig:xE}). The finite size effects for the microcanonical instability are more severe, as we now discuss. 

\subsubsection{Entropy barrier numerics}

\begin{figure}[t]
\includegraphics[width=\linewidth]{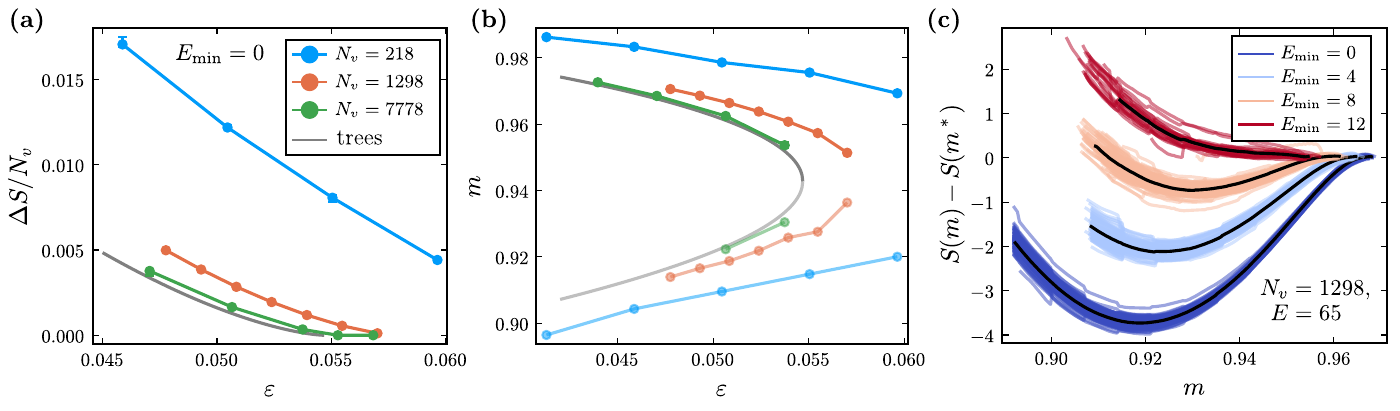}
\caption{\label{fig:entropy-barriers} (a) (Lower bound on) the height of the entropy density barrier in the valleys containing ground states, at energy density $\varepsilon$ for system size $N_v=218,1298,7778$ (girth 4,5,6 respectively). Error bars are smaller than the markers. Gray curve shows analytic calculation on trees ($N_v\rightarrow\infty$). (b) Average magnetization density at the local entropy maximum (top curves) and minimum (bottom light curves). Legend is the same as in (b). (c) Lower bound on the entropy relative to the maximum entropy within a valley, $S(m)-S(m^*)$, for system size $N_v=1298$, energy window $E=65\pm 1$ ($\varepsilon = 0.05$) and varying $E_{\rm min}$.  Light-colored curves are individual realizations, while the black curves show the average over samples (20 samples for $E_{\rm min}=12$ and $\approx 100$ for all remaining curves).}
\end{figure}

In the main text, we used the cavity method to determine the entropy landscape of ferromagnetic valleys (\autoref{fig:emin-escape}c) and estimated $\ec{micro}$ for higher-energy valleys from the maximum of $\ec{unstable}(x_0,x)$ (dashed curves) in~\autoref{fig:emin-escape}b. Now we compare those findings to numerics on finite-size closed graphs. \autoref{fig:entropy-barriers}a shows the magnitude of the entropy barrier (\autoref{eq:m-barrier}) as a function of energy density, averaged over 100 random realizations at each system size. The finite-size data alone may raise concern, as there is still a significant drift from $N_v=1298$ to $N_v=7778$. However, when compared to the tree calculation, the results are more encouraging, as $N_v=7778$ is close to converged. The same trend appears in \autoref{fig:entropy-barriers}b, which shows the average magnetization density at the entropy minimum and maximum alongside the analytically determined $m_1$ and $m_2$ (cf.~\autoref{fig:canonical-instability}b).

For higher-energy valleys, increasingly glassy dynamics with increasing $E_{\rm min}$ limited us to $N_v=1298$.~\autoref{fig:entropy-barriers}c shows the entropy landscape in the energy window $E\in[64,66]$, for $E_{\rm min}=0,4,8,12$. As $E_{\rm min}$ increases, the height of the barrier decreases, vanishing completely at $E_{\rm min}=12$. 

%% file: main-clean.bbl
\begin{thebibliography}{103}
\bibitem [{rrg()}]{rrg}%
  \BibitemOpen
  \href@noop {} {}\bibinfo {note} {A random regular graph is a graph where each vertex has the same degree, but the edges are otherwise assigned randomly. The girth is the length of the shortest cycle; as in Ref.~\cite{Placke_glass}, we adapt the method in Ref.~\cite{linial2019rhgrg} to construct RRGs with girth $\Theta(\log(N_v))$, where $N_v$ is the number of vertices.}\BibitemShut {Stop}%
\bibitem [{per()}]{perms}%
  \BibitemOpen
  \href@noop {} {}\bibinfo {note} {The local codespace for our chosen model is invariant under cyclic permutations, but not under arbitrary permutations, so we must fix some ordering of the spins around each node. On a tree, all orderings are equivalent by relabeling the branches. On a closed graph with loops, Ref.~\cite{Placke_glass} found that the arbitrary choice of labeling had no observable effect on the results.}\BibitemShut {Stop}%
\bibitem [{fp()}]{fp}%
  \BibitemOpen
  \href@noop {} {}\bibinfo {note} {When the superscript is omitted, $\Q{}$ should be interpreted as the fixed point distribution.}\BibitemShut {Stop}%
\end{thebibliography}
